\documentclass[aps, 10pt, prb, twocolumn, superscriptaddress, floatfix,
tighten]{revtex4-2}

\usepackage{xcolor} \usepackage{amsmath} \usepackage{mathrsfs}
\usepackage{amssymb} \usepackage{bm} 
\usepackage{graphicx} \usepackage[unicode=true,pdfusetitle,
bookmarks=false,bookmarksnumbered=false,bookmarksopen=false,
breaklinks=false,pdfborder={0 0
  1},backref=false,colorlinks=true,urlcolor=blue,citecolor=blue,linkcolor=black]
{hyperref} \usepackage{subfigure} \usepackage{textcomp}
\usepackage{microtype} 
\usepackage{braket}

\newcommand{\ord}[1]{{\mathit{O}}\left(#1\right)}

\newcommand{\bsub}{\begin{subequations}}
  \newcommand{\esub}{\end{subequations}}

\newcommand{\vex}[1]{\bm{\mathrm{#1}}} \newcommand{\e}{\epsilon}
\newcommand{\pd}{\partial} 
 
\newcommand{\transduction}{T} \newcommand{\transmission}{P}
\newcommand{\damping}{\gamma} \newcommand{\skyrmsize}{\lambda}
\newcommand{\lambdaE}{Q\alpha} \newcommand{\Lspace}{l_\perp}

\begin{document}

\title{Magnon scattering and transduction in Coulomb-coupled quantum
  Hall ferromagnets }

\def\rice{Department of Physics and Astronomy, Rice University,
  Houston, Texas 77005, USA} \def\rcqm{Rice Center for Quantum
  Materials, Rice University, Houston, Texas 77005, USA}

\author{Alexander Canright}\affiliation{\rice} \author{Deepak Iyer}
\affiliation{Department of Physics \& Astronomy, Bucknell University,
  Lewisburg, Pennsylvania 17837, USA} \author{Matthew
  S. Foster}\affiliation{\rice}\affiliation{\rcqm}

\date{\today}

\begin{abstract}
  The magnetization field of a quantum Hall ferromagnet (QHFM) can
  host a variety of spin textures, including skyrmions and
  magnons. When projected into the lowest Landau level with $\nu=1$
  filling, the topological (Pontryagin) charge density of the
  magnetization field is proportional to the electric charge density,
  allowing for long-range spin-spin interactions.  Inspired by recent
  experimental developments that enable all-electrical generation and
  detection of magnons, in this work we theoretically demonstrate
  two phenomena that can occur due to Coulomb interactions that are
  unique to QHFMs: magnons can scatter off of point charges at a
  distance, and skyrmions can act as \emph{transmitters} and
  \emph{receivers} for magnons to be transduced between separate
  layers of a bilayer QHFM. The latter Coulomb-mediated spin drag
  effect occurs at arbitrary distance and could facilitate long-range
  magnonics, such as detection of spin waves for future experiments in
  2D materials.

\end{abstract}

\maketitle

\def\verticaldistance{15pt}

\section{Introduction}

Quantum Hall ferromagnets (QHFMs) \cite{girvinQuantumHallEffect1999}
have come once again to the forefront of condensed matter research,
thanks to experimental developments in 2D quantum materials. Due to
the flat-band dispersion of the (idealized) Landau level, QHFMs
conceptually serve as the simplest type of itinerant
magnet (with a fully spin polarized, unentangled product ground state), but
also provide a more general paradigm for interaction-induced
spontaneous symmetry breaking in topological flat-band materials. In
addition, QHFMs feature strong locking between charge and spin degrees
of freedom, so that topological skyrmion spin configurations \cite{girvinQuantumHallEffect1999,Sondhi93,Fertig94,Fertig97} and
propagating magnons \cite{bychkovTWODIMENSIONALELECTRONSSTRONG1981,Kallin84} carry electric monopole and dipole moments,
respectively
\cite{gorkovContributionTheoryMott1968,Lerner78}.

Recent attention has focused on elucidating the QHFM ground states and
excitations in multivalley and/or multilayer graphene systems.
Studies have addressed, for example, the realization and stability of skyrmions in
low-energy Dirac Landau levels
\cite{barrettOpticallyPumpedNMR1995,Cote97,Sinova00,Yang06,Lian17,jolicoeurQuantumHallSkyrmions2019,Zhou20,Atteia21,Pierce22},
as well as the nature of the $\nu = 0$ insulating state of monolayer
graphene
\cite{Abanin06,kharitonov,Young12,Young14,jolicoeurQuantumHallSkyrmions2019,Liu22}.
Another flurry of activity was triggered by the discovery of
correlated states \cite{Cao18ins} and superconductivity
\cite{Cao18sc,Yankowitz19,Lu20} in flattened bands of twisted bilayer
graphene (TBLG) \cite{MacDonald19}. These nearly flat bands can host
Chern numbers \cite{Balents20,Ledwith21} and insulating QHFM analogs
\cite{Cao18ins,Sharpe19,Lu20}. The observation of superconductivity in
TBLG has supercharged the field of moir\'e materials, with some theoretical
proposals linking this to the condensation of QHFM skyrmion pairs
\cite{Khalaf21,Ledwith21}.

Yet another important recent experiment demonstrated all-electrical
generation and detection of magnons in the $\nu = 1$ state of
monolayer graphene \cite{weiElectricalGenerationDetection2018}. Here,
charge is injected between contact-induced \emph{non-sample-spanning}
$\nu = 2$ edge states and the $\nu = 1$ edge states that connect the
source and drain contacts. For a bias exceeding the Zeeman energy gap,
electrons tunnel between the $\nu = 2$ and $\nu = 1$ edges. This
interedge tunneling induces a spin flip, emitting a magnon into the
device (see Figure~1 in Ref.~\cite{weiElectricalGenerationDetection2018}
for an illustration). The magnons are electrically detected via the
suppression of the Hall conductance from its quantized value, because
magnon emission and absorption serves as a source of remote inelastic
scattering between distant chiral edge states. This mechanism is distinct from established optical methods of magnon generation \cite{pinczukSpectroscopicMeasurementLarge1992}. Additional experiments
exhibited similar phenomena in a quantum-Hall antiferromagnet
\cite{Stepanov18}.  Strong damping of magnon transmission at higher
dopings away from $\nu = 1$ was taken as evidence for the formation of
a Skyrme crystal, whose phonons can dissipate the magnon energy
\cite{Zhou20,Cote97}, see also Ref.~\cite{Chakraborty23}. The electric dipole moment carried by magnons
was probed via a related setup in \cite{Assouline21}. Nonequilibrium
magnon-skyrmion dressing was investigated in \cite{Pierce22}.

In this work, we theoretically explore novel magnonic phenomena
enabled by electrical magnon generation, and facilitated by the strong
spin-charge locking \emph{unique} to QHFM devices. We simulate magnon
injection into a spin SU(2) QHFM using semiclassical spin dynamics.
We show that external electric charges deflect magnons, evidently due to the
effective electric dipole moment carried by the latter. We further compare the
results of the spin semiclassics to a second-order Born approximation
for the effective magnonic Schr\"odinger
equation, and find that these give similar results.

\begin{figure}[t]
  \centering
  \includegraphics[width=0.44\textwidth]{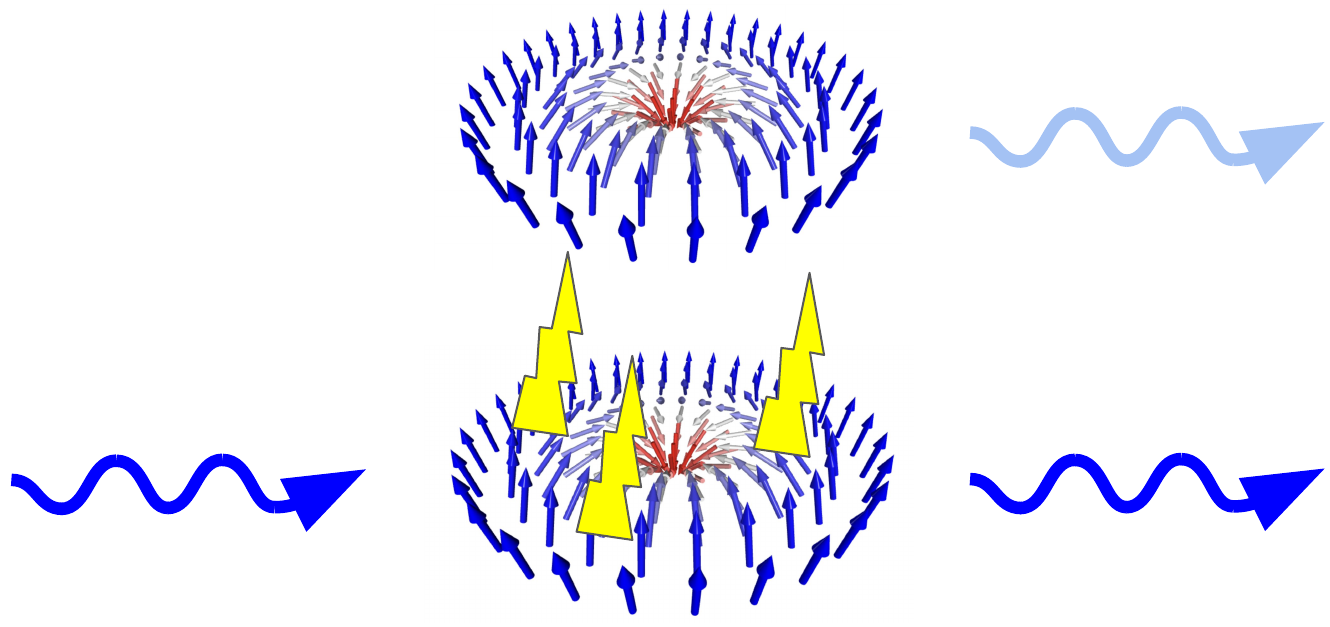}
  \caption{Cartoon of Coulomb-mediated magnon-skyrmion-skyrmion-magnon
    transduction.  Magnons injected in the lower layer induce
    undulations of the skyrmion core in that layer.  These undulations
    couple via the Coulomb interaction to the noncollinear spin
    texture in the upper layer (associated e.g.~to another proximal
    skyrmion). The induced core fluctuations in the top-layer core
    then function as an antenna, emitting directed magnon radiation
    into the upper layer.}
  \label{fig:transduction_cartoon}
\end{figure}

Second, we demonstrate a novel type of Coulomb-mediated \emph{spin
  drag} that is directly enabled by skyrmion defects expected to be
present near $\nu = 1$
\cite{barrettOpticallyPumpedNMR1995,Yang06,Lian17,jolicoeurQuantumHallSkyrmions2019}. We consider a
bilayer system of two QHFMs separated by an insulating barrier. We
assume negligible tunneling, but take into account the Coulomb
interactions between the spin textures in the two layers. We focus
upon
a geometry with a pair of skyrmions co-located in the top and bottom
layers, pinned by an impurity potential. We show that magnons injected
into the bottom layer are \emph{transduced} into the top layer. The
mechanism involves core undulations of the Coulomb-coupled skyrmions in
the two layers, such that the top-layer skyrmion functions as a magnon
emitter, see Figure~\ref{fig:transduction_cartoon} for an illustration
of the setup.  We show that experimentally realistic parameters should produce a
detectable effect.

\subsection{Outline}

The rest of this paper is organized as follows.  In
Sec.~\ref{sec:MainResults} we define the model and describe the
simulation method.  Then we present our two main physics results,
consisting of magnon-Coulomb charge scattering in a monolayer, as well
as Coulomb-mediated magnon transduction (spin drag) between two
monolayers separated by an insulating barrier.  In Sec.~\ref{sec:MP}
we provide details of the numerical implementation for the dynamics
and observables.  In Sec.~\ref{sec:SLDyn}, we consider single-layer
magnon dynamics in more detail.  In addition to magnon-charge
scattering, we also benchmark our numerical methods by considering
skyrmion-magnon scattering due to stiffness alone, previously
considered in the context of ordinary chiral ferromagnets
\cite{garst,iwasakiTheoryMagnonskyrmionScattering2014,liTowardsQuantumMagnonics2022,jiangSkyrmionsMagneticMultilayers2017}.
We consider future directions and conclude in Sec.~\ref{sec:Conc}.

\section{Main results \label{sec:MainResults}}

\subsection{Model and equations of motion}

An SU(2) QHFM in the lowest Landau level (LLL) at filling factor
$\nu = 1$ has the effective real-time action
\cite{girvinQuantumHallEffect1999}
\begin{subequations}
  \begin{align}
    \!\!\!
    S_M&=
         \int d t \, d^2\vex{r}
         \left[ -s\,n\,\dot{m}^i\mathscr{A}_i(\vec{m}) + \lambda (m^im^i-1)\right] \\
       &+ 
         \int d t \, d^2\vex{r}
         \left[s\, n\, g\,\mu \, B^i m^i -\frac{\rho_s}{2}\pd_\alpha m^i\pd_\alpha m^i\right] 
         \label{QHMF_normal}
  \end{align}
  \begin{align}
       &+
         \int d t \, d^2\vex{r}
         \left\{
         \begin{aligned}
           &\, e\varrho(\vec{m})A^0
           \\&\, - \frac{e^2}{2\e} \int d^2\vex{r'}
           \frac{\varrho[\vec{m}(t,\mathbf{r})]\varrho[\vec{m}(t,\mathbf{r'})]}{|\mathbf{r-r'}|}
         \end{aligned}
    \right\}\!.
    \label{QHMF_action_electro} 
  \end{align}
  \label{QHMF_action}
\end{subequations}
This is a ferromagnetic O(3)/O(2) nonlinear sigma model with
magnetization field variable $\vec{m}(t,\mathbf{r})$. The Berry phase
term $\dot{m}^i\mathscr{A}_i(\vec{m})$ encodes the non-inertial
dynamics of the spins, while $\lambda(t,\vex{r})$ is a Lagrange
multiplier field enforcing the constraint $\vec{m}\cdot\vec{m} = 1$.
Here $s = 1/2$ and $n={1}/{2\pi l_B^2}$ is the average electron
density, with $l_B = \sqrt{\hbar c/e B}$ the magnetic length. In
Eq.~(\ref{QHMF_normal}), the Zeeman coupling to the external magnetic
field $B^i$ is via $g \mu$, the Bohr magneton times the electron
$g$-factor, while $\rho_s$ denotes the spin stiffness coefficient
(with units of energy).  In Eq.~(\ref{QHMF_action_electro}), $e>0$ is
the magnitude of the electron charge, $A^0(t,\mathbf{r})$ is the
scalar electric potential, and $\e$ the dielectric constant of the
material.  Due to the LLL projection, the spin texture couples to
electric charge through the topological Pontryagin density
\begin{equation}\label{Pontryagin}
  \varrho(\vec{m})
  \equiv
  \frac{1}{4\pi}\vec{m}\cdot\left(\pd_x \vec{m}\times\pd_y \vec{m}\right).
\end{equation} 
We denote 3D vectors with an overarrow and their components with Latin
indices, and 2D vectors using boldface with components having Greek
indices, e.g.  $\vec{m}(t,\mathbf{r})$ and
$\partial_\mu m^i(t,\mathbf{r})$ ($\mu \in \{x,y\}$,
$i \in \{1,2,3\}$).

\begin{figure*}[th!]
  \subfigure[]{ \centering
    \includegraphics[width=0.315\textwidth]{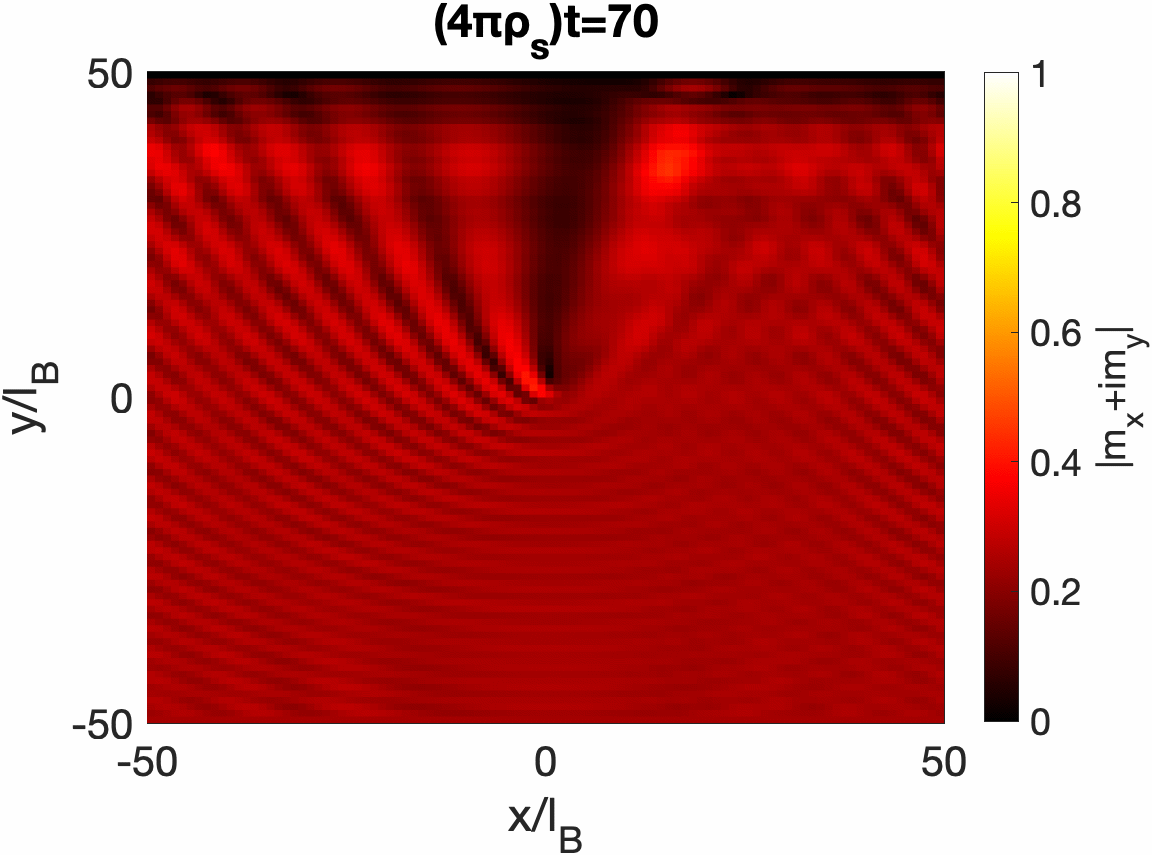}
  } \subfigure[]{ \centering
    \includegraphics[width=0.315\textwidth]{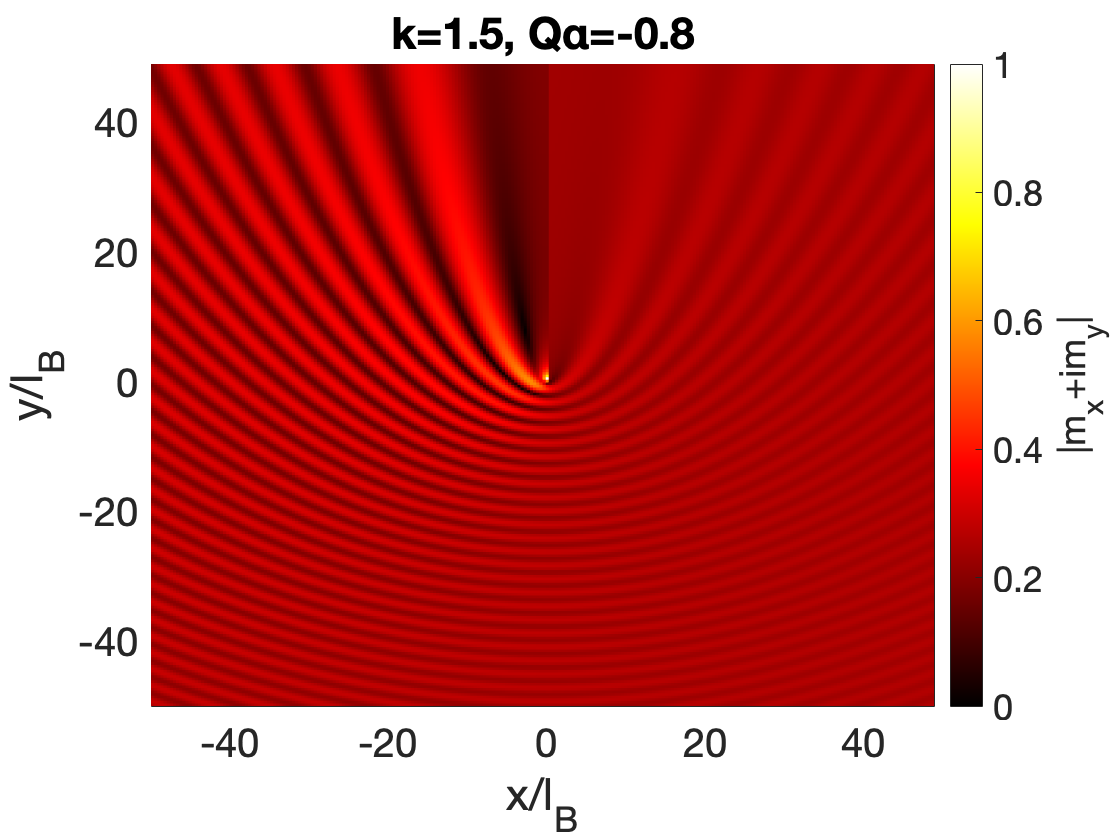} }
  \subfigure[]{
    \includegraphics[width=0.315\textwidth]{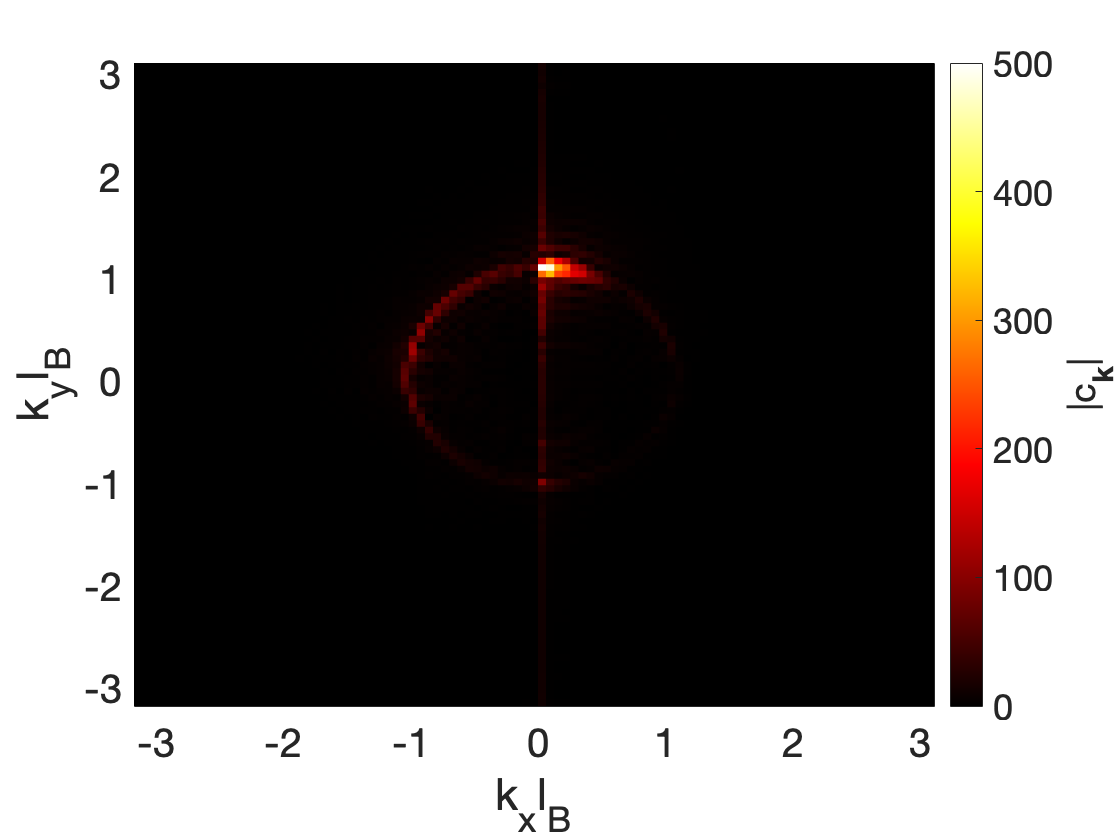}
  } \hspace{1mm}
  \caption{(Color online) Plane magnons scatter off of the electrical field due to a
    point charge located at $(x,y,z) = (0,0,\Lspace)$.  Despite
    carrying zero net electric charge, magnons interact with electric
    fields due to the effective electrical dipole moment proportional
    to the U(1) spin current, Eq.~(\ref{dipole}).  (a) Snapshot of a
    numerical simulation. The color indicates the magnitude of the
    deviation of the spin field from the ferromagnetic ground state
    $|\psi(x,y)|\equiv\sqrt{m_x^2+m_y^2}$ after plane magnons are
    driven into the sample from the bottom. The scattering arises due
    to the electric field of a point charge placed at a distance
    $\Lspace = 2$ above the plane. The magnon has wavevector
    $k=1.5708$ and the magnitude of the point charge is $Q=-9$.  Here
    all distances and inverse wavenumbers are measured in units of the
    magnetic length $l_B = 1$, and charge in units of the electron
    charge $e > 0$.  (b) Analytical result via the second Born
    approximation for the magnon scattering off of a point charge at
    $(x,y,z)=(0,0,0)$, with $k=1.5$ and $\lambdaE=-0.8$, where
    $\alpha$ is defined via Eq.~(\ref{dimparams}).  (c) $k$-space
    distribution of magnons at end of simulation. Here, $c_\mathbf{k}=\sum\limits_{\mathbf{r}}e^{-2\pi i\mathbf{k\cdot r}/N}\psi(\mathbf{r})$ is the discrete Fourier transform for the mode $\mathbf{k}$ of the magnon field $\psi=m_x+im_y$.}
\label{fig:main-results-pointcharge}
\end{figure*}

From this, we can derive the semiclassical equations of motion for the
magnetization field:
\begin{subequations}
  \begin{align}
    \frac{d\vec{m}}{dt}
    &= \vec{m}\times\vec{\mathcal{B}}_\text{eff}, \\
    \vec{\mathcal{B}}_\text{eff} 
    &= g\mu B\hat{z} + \frac{\rho_s}{sn}\nabla^2\vec{m} \\
    - & \frac{e}{4\pi sn}
        [\e^{\mu\nu}\vec{m}\times\pd_\nu\vec{m}]E^{ext}_\mu   
        \nonumber\\
    + & \frac{e}{4\pi sn}
        [\e^{\mu\nu}\vec{m}\times\pd_\nu\vec{m}]
        \left\{\frac{e}{\e} \int d^2\vex{r'}
        \frac{(\mathbf{r}-\mathbf{r}')_\mu}{|\mathbf{r}-\mathbf{r}'|^3} 
        \varrho[\vec{m}(t,\mathbf{r}')] 
        \right\}, 
  \end{align}
\label{QHMF_eom}
\end{subequations}
where $E^{ext}_\mu = - \partial_\mu A^0(t,\vex{r})$ is the in-plane
external electric field and $\e^{\mu\nu}$ is the Levi-Civita tensor.

We solve the semiclassical equations (\ref{QHMF_eom}) via 4th-order
Runge-Kutta method for spins in a finite $L \times L$ sample, with lattice
spacing set by $l_B$. Forced and absorbing boundary conditions are
employed to inject and ``detect'' magnons, respectively.  In the rest
of this section, we describe the main setup and results of this work,
including estimates for dependence upon certain parameters relevant to
experimental detection of the predicted effects.

\begin{figure*}[t!]
  \subfigure[]{ \centering
    \includegraphics[width=0.31\textwidth]{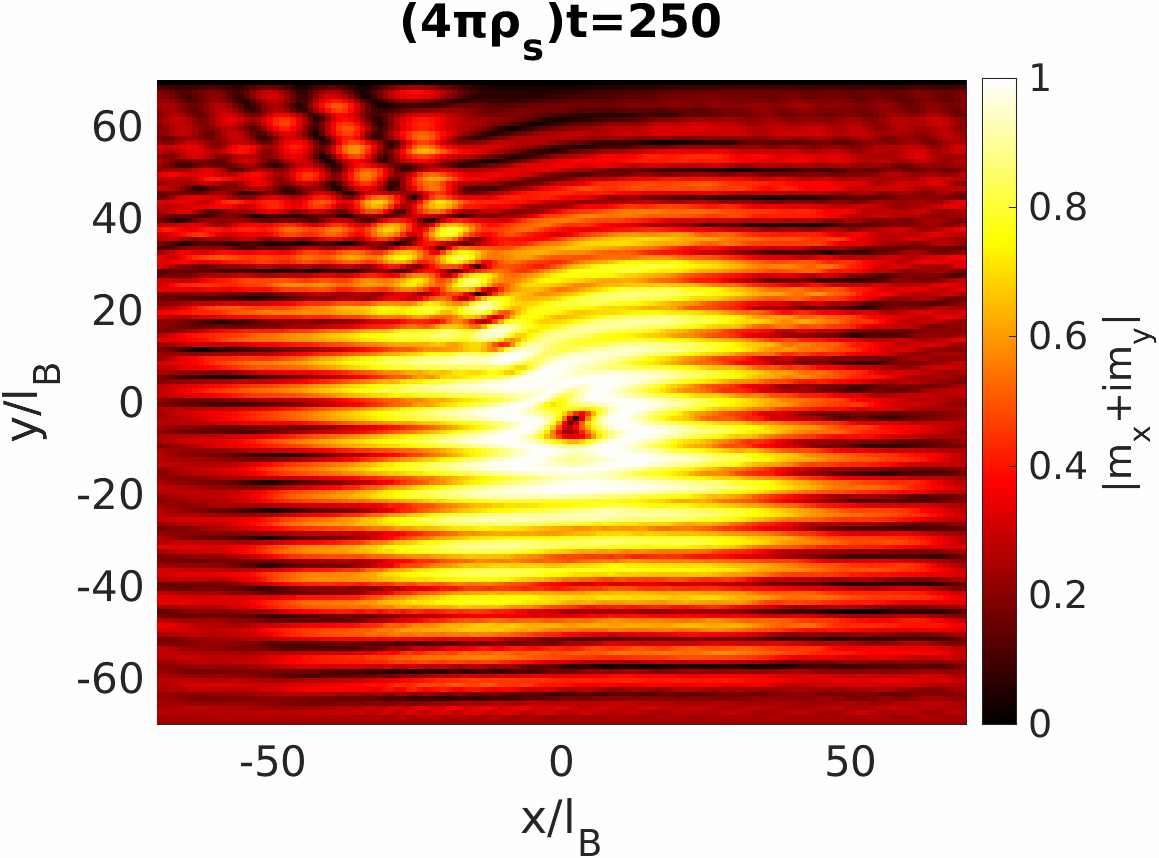}
  } \subfigure[]{ \centering
    \includegraphics[width=0.31\textwidth]{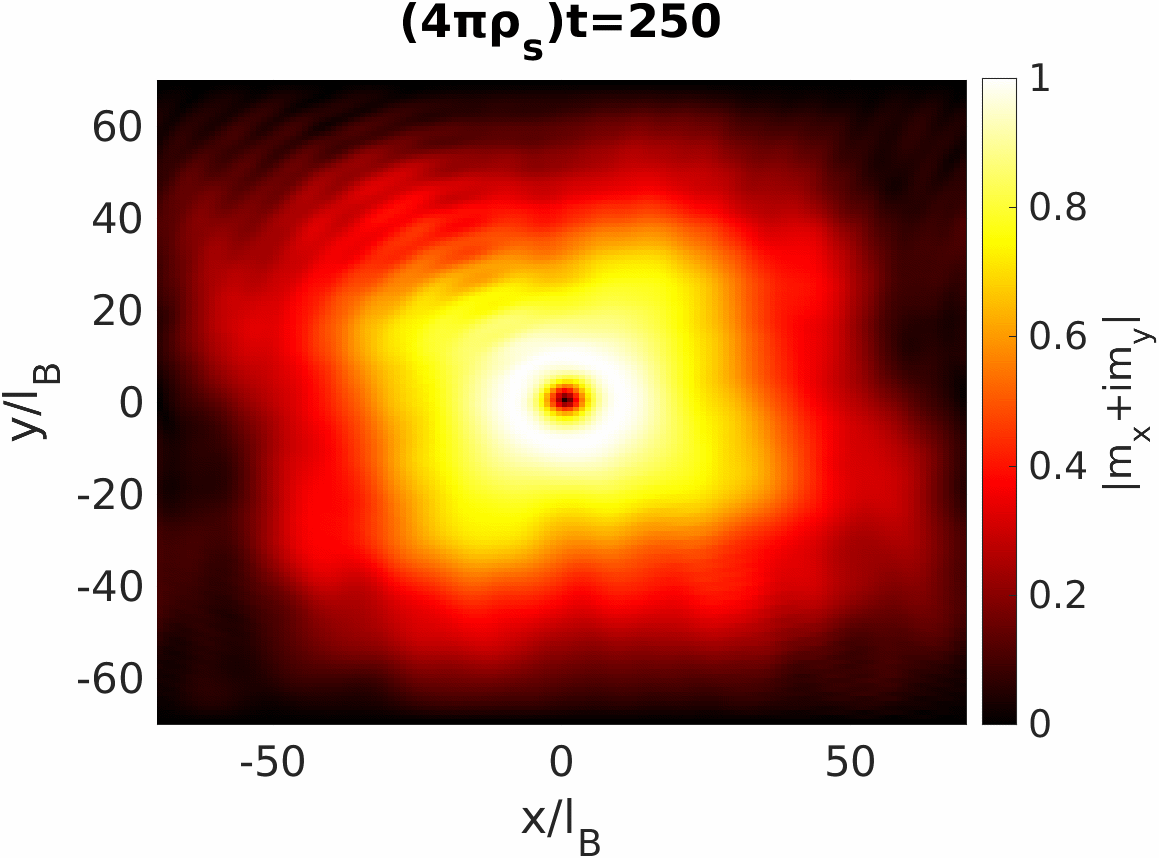}
  } \subfigure[]{ \centering
    \includegraphics[width=0.31\textwidth]{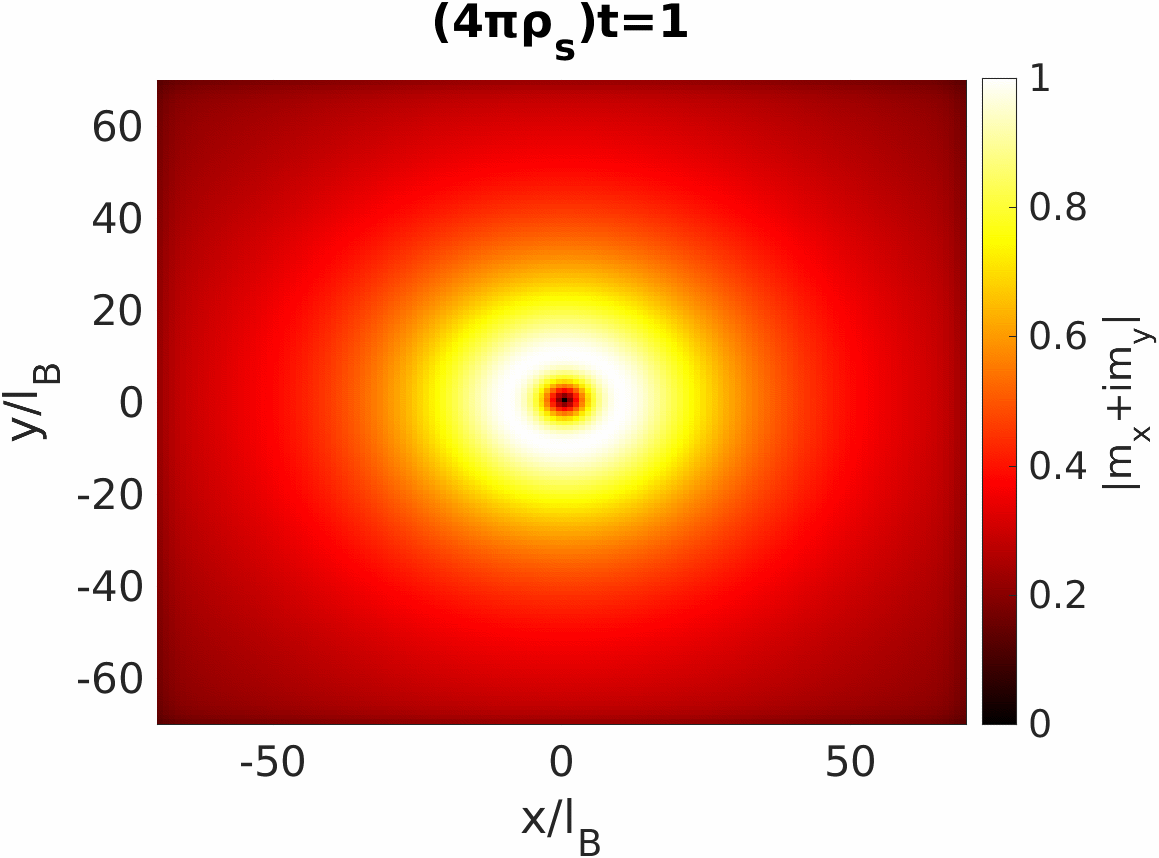}
  } \hspace{1mm}
  \caption{Interlayer magnon transduction (``spin drag'') due to
    Coulomb-mediated skyrmion-skyrmion interactions.  A cartoon of the
    setup is shown in Figure~\ref{fig:transduction_cartoon}. We simulate
    two QHFMs separated by small distance of $0.1 \,
    l_B$, 
    each with a skyrmion defect of size $\skyrmsize=5\,l_B$. The skyrmion size $\skyrmsize$, defined in Appendix A 
    [Eq.~\eqref{skyrmionsize}], is half of its radius. The defects are vertically stacked
    and pinned by an impurity potential. Plane-wave magnons are driven
    from the bottom of the spin field in layer 1. Magnon-skyrmion
    interactions in that layer produce a dynamic undulation in the
    skyrmion core, and this acts as a ``transmitter'' that (via
    Coulomb interaction) drives core undulations in the layer-2
    skyrmion (the ``receiver''). The latter produces directed magnon
    emission in the second layer.  (a) Magnitude of the lateral
    magnetization field $\sqrt{m_x^2+m_y^2}$ in layer 1, in which
    magnons are driven from the bottom.  (b) Magnitude of the lateral
    magnetization in layer 2, in which magnons can be seen emitted
    from the skyrmion, towards the upper-left of the magnetization
    field. Plots (a,b) are taken within a steady state at a late time
    in the simulation $\tau = 250$.  Note that the anisotropy of the
    transduced spin texture (b) clearly reflects the magnon-skyrmion
    scattering occuring in the driven layer (a).  (c) Initial stable
    skyrmion texture in each layer. All results have the
    \emph{interlayer} ``interaction strength'' $\alpha_{12} = 1$
    [Eq.~(\ref{dimparams})].}
  \label{fig:main-results-transduction}
\end{figure*}

Our numerical results show evidence of two novel phenomena in LLL
QHFMs: magnon electrodynamics in ferromagnets, and
Coulomb-facilitated, skyrmion-mediated spin-drag between layers. In
the first case, the system is initialized to a ferromagnetic texture
and a point charge placed near the center but above the plane of the
QHFM.  Plane magnons are driven up from the bottom row and collected
by an absorbing boundary layer in the top rows. The magnons scatter
from the electric field despite carrying no net electrical charge, see
Figure~\ref{fig:main-results-pointcharge}.  This interaction occurs due
to the \emph{dipole moment} of the propagating magnons, proportional
but perpendicular to the conserved U(1) spin current
\cite{gorkovContributionTheoryMott1968,Lerner78,Kallin84}.  In the second case, the system is extended
to two layers that are coupled via the Coulomb interaction, each
initialized with a stable skyrmion texture, and magnons are driven
across the skyrmion in one of the layers. We quantify the ``spin
drag'' due to the Coulomb-mediated \emph{transduction} of magnon
energy from the driven layer to the ``receiver'' layer, see
Figures~\ref{fig:main-results-transduction} and
\ref{fig:transduction_hero_graphs}.

In order to discuss these results in more detail, we first summarize
the key parameters for the calculations.  After discretizing the spin
equations of motion (EOM) in space ($N\times N$ lattice sites,
$N = L/l_B$) and time, we recast them in terms of a dimensionless time
variable $\tau$ and three dimensionless parameters,
\bsub\label{dimparams}
\begin{gather}
  \tau \equiv (4\pi\rho_s)t, \qquad b \equiv
  \left(\frac{1}{4\pi\rho_s}\right)g\mu B,
  \\
  \boldsymbol{\mathcal{E}} \equiv
  \left(\frac{1}{4\pi\rho_s}\right)el_B\boldsymbol{E}^{ext}, \qquad
  \alpha \equiv \left(\frac{1}{4\pi\rho_s}\right)\frac{e^2}{\e l_B}.
\end{gather}
\esub Here $b$ and $\boldsymbol{\mathcal{E}}$ denote the dimensionless
out-of-plane Zeeman and in-plane external electric fields,
respectively.  The stiffness determines the timescale, so it can be
tuned by scaling $\tau$.  Physically, the spin stiffness arises from
Coulomb exchange and takes a value $\rho_s \propto e^2/\epsilon l_B$,
which is the same energy scale for the residual Coulomb interaction at
integer LLL filling \cite{bychkovTWODIMENSIONALELECTRONSSTRONG1981,Kallin84}. Therefore the interaction parameter $\alpha$ is
actually an order-one number, independent of $e$ and $l_B$.
Eq.~(\ref{discrete_EOM}) details the implementation of these
parameters in the discretized EOM.

\subsection{Magnon-Coulomb charge scattering}

Figure~\ref{fig:main-results-pointcharge} shows the results of a
simulation in which magnons interact with a point charge.  In a
ferromagnetic texture with no skyrmion, plane magnons with topological
charge density $\varrho(\vec{m})=0$ everywhere are driven from the
bottom row and subjected to an electric field due to a point charge at
the origin, held at a distance of $\Lspace = 2$ magnetic lengths above
the plane in the $z$-direction.

The injected magnons scatter off of the point charge, leading to the
diffraction pattern shown in
Figure~\ref{fig:main-results-pointcharge}(a).  While non-topological
spin textures carry zero electric charge, the LLL projection endows
magnons with an effective dipole moment \cite{gorkovContributionTheoryMott1968,Lerner78,Kallin84},
\begin{align}\label{dipole}
  \vec{d} = \frac{e}{4 \pi} \vec{J} \times \hat{z},
\end{align}
where $\vec{J}$ denotes the U(1) spin current, see
Eq.~(\ref{spincurrent}).

In this simulation, the initial texture is
$\vec{m}(t=0,\mathbf{r})=\hat{z}$, a pure ferromagnet.  The parameters
are $N=100$ and dimensionless Zeeman strength $b=0.13$, while the
magnon carries wavenumber $k_y=1.5708$.  Unless otherwise noted, we
set the lattice spacing (magnetic length) $l_B = 1$ and measure
charges in units of $e$.  The scalar electric potential is due to an
unscreened point charge placed a distance $\Lspace = 2$ above the
plane of the sample, with charge $Q = - 9$.
We drive the bottom row of spins to precess about the $\hat{z}$-axis
with frequency $\Omega = b + 2.0$, so as to induce a plane spin wave
that traverses the sample from bottom to top.  [Dimensionless
frequency is measured with respect to the stiffness energy, defined in
terms of the dimensionless time $\tau$: $\Omega = 2 \pi / \tau$, see
Eq.~(\ref{dimparams}).]
The stiffness (Laplacian) term in Eq.~(\ref{QHMF_eom}) has periodic
boundary conditions in the $x$-direction so that the walls do not
reflect the outgoing spin wave profile. Along the top edges in each
layer, we add an absorbing boundary layer with an exponential damping
profile so as to catch and dissipate the magnons with minimal
reflection \cite{abl}. The damping itself is implemented via a Gilbert
damping term in the equation of motion, see Eq.~\eqref{LLG}. Here, the
damping profile is set to $\damping(\mathbf{r})=\damping_0\exp[ -(y-N)/2]$,
with $\damping_0=1.6$. 

For weak point charges, the outgoing waves appear mostly symmetrically
distributed, but for stronger point charges, circular waves can be
seen emitted towards the left or right, depending on the sign of the
charge (negative or positive, respectively).

In the
continuum, QHFM magnons can be interpreted as number-conserving,
non-relativistic quantum particles subject to \emph{synthetic} scalar
and vector potentials due to the physical electric field
$\boldsymbol{\mathcal{E}}$ [Eq.~(\ref{dimparams})]. The effective
single-particle Hamiltonian is
\begin{subequations}
  \begin{align}
    \hat{h} 
    &= 
      b + \left[\mathbf{\hat{P}}-\boldsymbol{\mathcal{A}}(\hat{\mathbf{r}})\right]^2 + \mathcal{V}(\mathbf{\hat{r}}), 
    \\
    \mathcal{A}^\beta(\mathbf{r}) 
    &\equiv 
      -\frac{\e^{\beta\alpha}}{2}\mathcal{E}^\alpha(\mathbf{r}), 
    \\
    \mathcal{V}(\mathbf{r}) 
    & \equiv 
      -\boldsymbol{\mathcal{A}}^2(\mathbf{r}).
  \end{align}\label{SyntheticPotentials}
\end{subequations}
Using the second Born approximation for plane-magnon scattering with
this Hamiltonian yields Figure~\ref{fig:main-results-pointcharge}(b),
which closely resembles the real-space pattern of our numerical
results. Its angular form factor also resembles the $k$-space
distribution shown in Figure \ref{fig:main-results-pointcharge}(c).  We
note that the numerical and analytical results shown in
Figs.~\ref{fig:main-results-pointcharge}(a,b) are not expected to
produce identical results, because the second Born approximation is
simple to evaluate only for a point charge located in the plane of the
magnet, whereas the simulation has $\Lspace = 2$ (to prevent a
short-range divergence due to the Coulomb potential).
Our simulation techniques also produce reliable results for
magnon-skyrmion scattering, which can be used to benchmark our code
and compare to existing findings, see Figure~\ref{fig:scattering_graphs}. We discuss these results in detail in
Section~\ref{sec:SLDyn}.

We also observe qualitatively similar Coulombic magnon scattering when
simulating a bilayer of two QHFMs, separated by a thin insulating
layer that prevents tunneling between the layers but facilitates
interlayer Coulomb interactions.  When magnons are driven in one layer
in the ferromagnetic ground state, and the second layer contains a
single skyrmion, magnons in the first layer scatter off of the
skyrmion at a distance due to the electric field it
produces. Therefore, to experimentally verify magnon electrodynamics,
one can either apply a nonuniform external electric field or simply
dope a nearby layer with electrons or holes.  The dipolar character of
magnons in QHFMs was recently confirmed by interferometric
measurements \cite{Assouline21}.

\subsection{Spin drag via Coulomb-mediated magnon-skyrmion
  transduction}
\begin{figure*}[ht!]
  \subfigure[]{
    \includegraphics[width=0.31\textwidth]{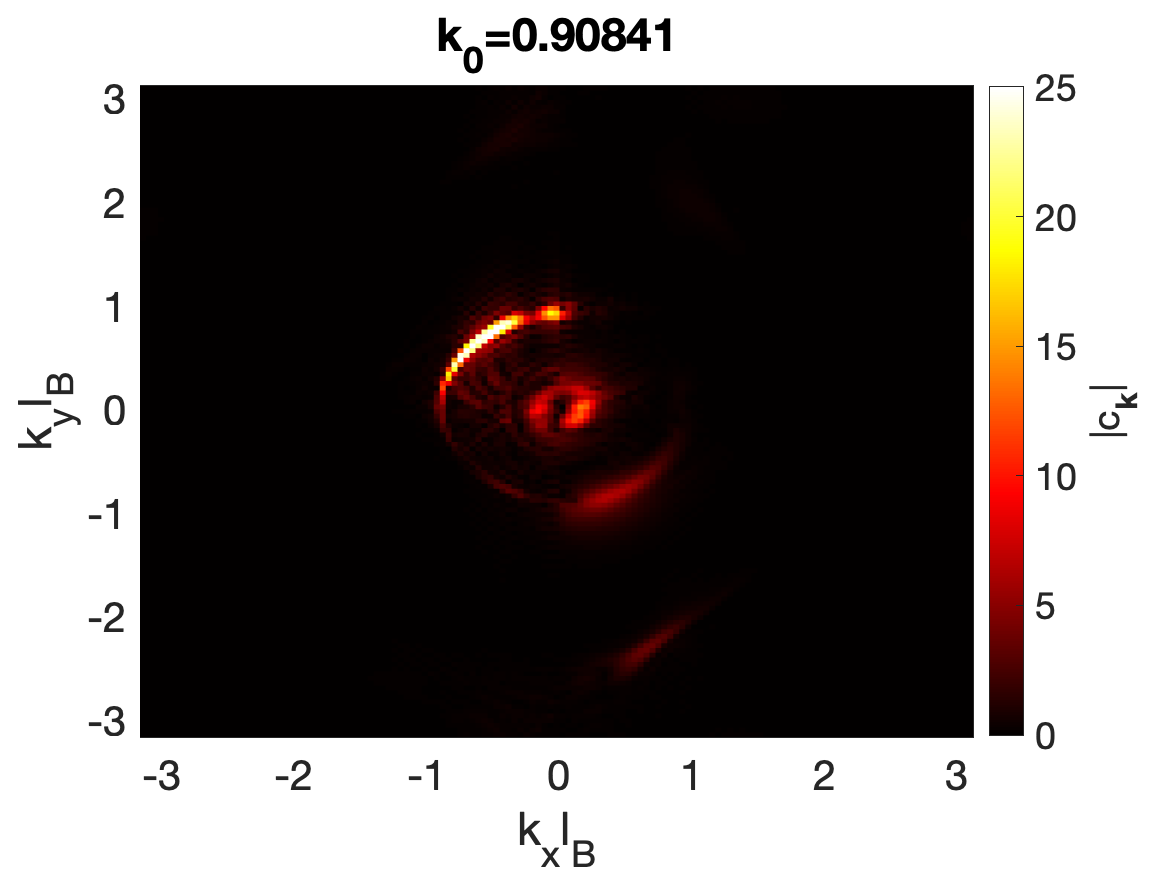}
  } \subfigure[]{ \centering
    \includegraphics[width=0.31\textwidth]{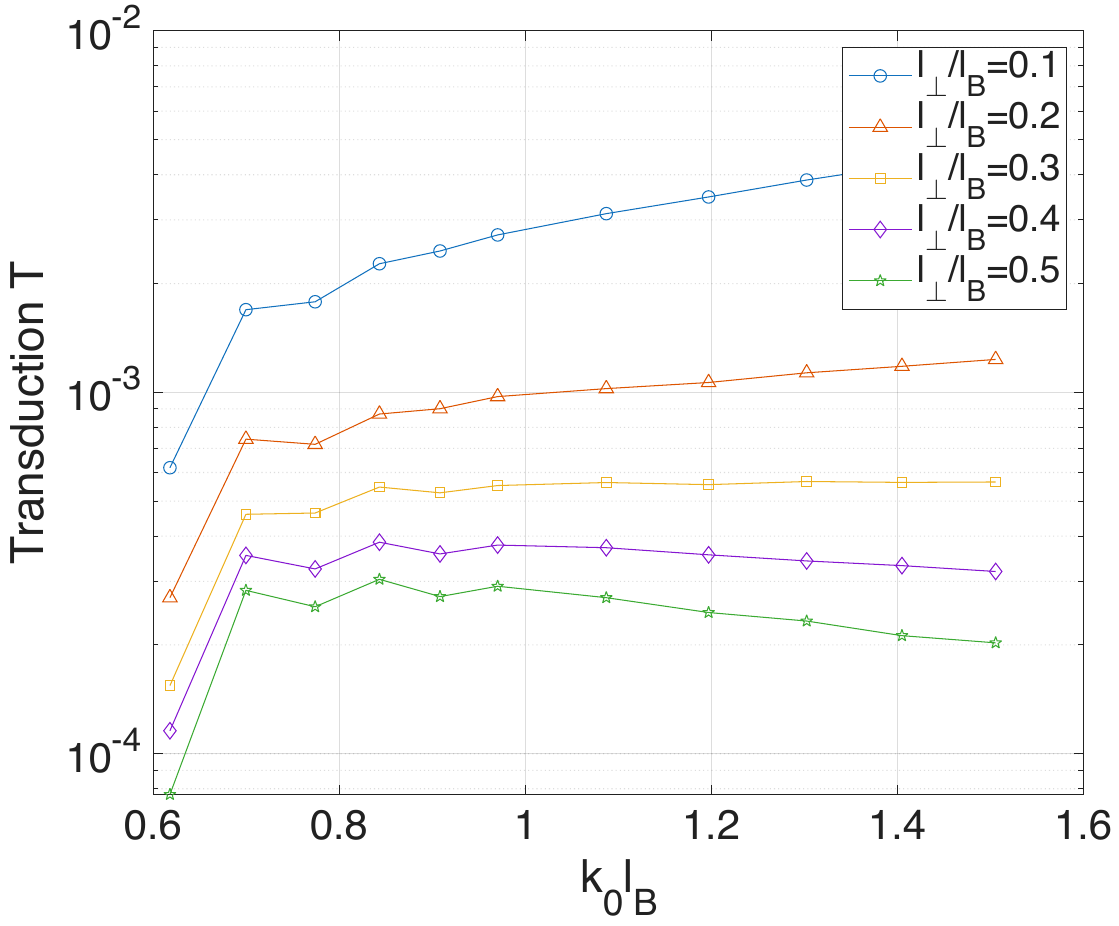}
  } \subfigure[]{
    \includegraphics[width=0.31\textwidth]{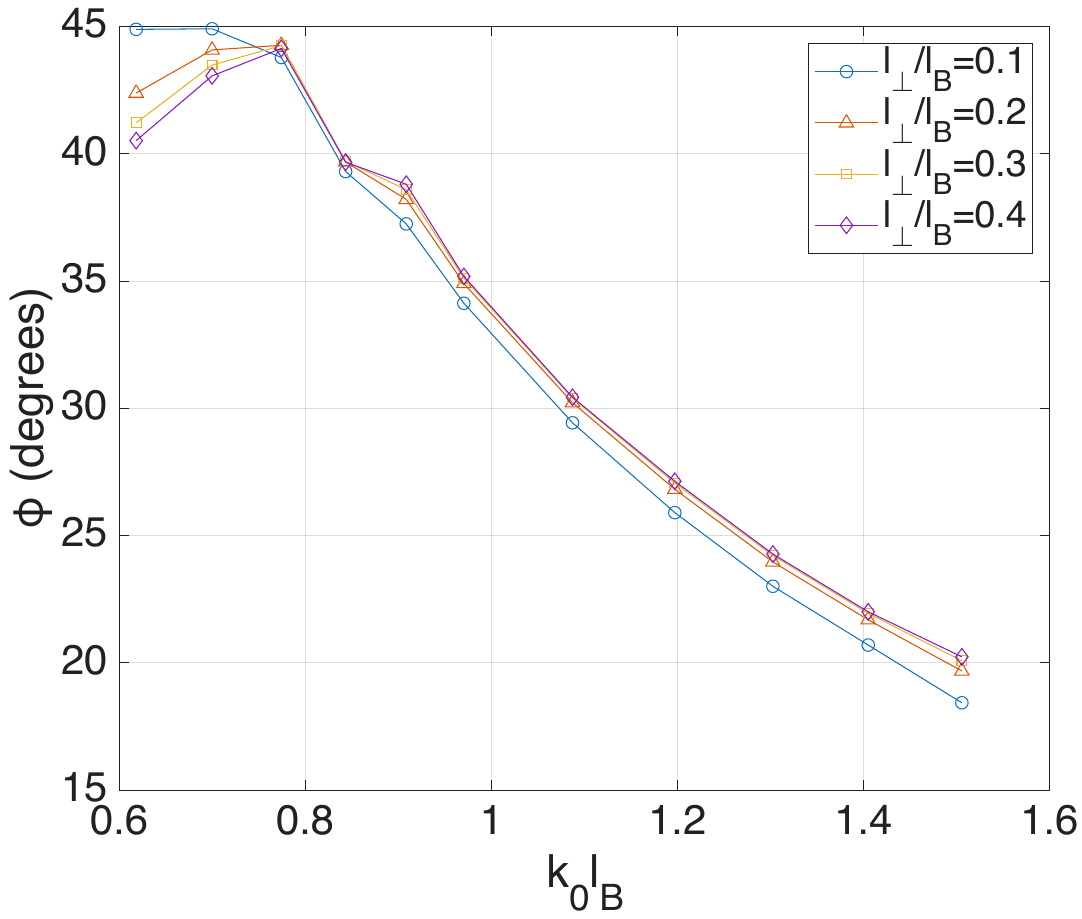}
  } \\ \hspace{1mm}
  \caption{Parameter dependence of Coulomb-mediated spin-drag
    simulations I.  (a) Magnitude of the difference in the $k$-space
    distribution of magnons in the receiving layer at the end of a
    driven simulation versus undriven simulation.  The arc shows that
    transduced magnons in the second layer share the wavenumber
    $k_0\approx0.90841 \, l_B^{-1}$ of the skyrmion-scattered magnons
    in the driving layer (not shown).  (b) Transduction ratio $T$
    [Eq.~(\ref{transductionRT})] versus driven magnon wavenumber
    $k_0$ and interlayer spacing $\Lspace$ for various simulations,
    with skyrmion size $\skyrmsize=5\,l_B$, interaction parameter $\alpha=1$,
    and Gaussian pinning potential in each layer. Note that the
    $k$-values deviate by about $5\%$ from the continuum quadratic
    dispersion for high $k$, since they obey the lattice magnon
    dispersion relation Eq.~\eqref{dispersion}.  (c) Average
    magnong scattering angle $\phi$ in layer 2 versus driven magnon wavenumber
    $k_0$ and interlayer spacing $\Lspace$ for various simulations,
    with skyrmion size $\skyrmsize=5\,l_B$, interaction parameter $\alpha=1$,
    and Gaussian pinning potential in each layer. The scattering
    profile of transduced magnons always resembles that of the driving
    layer, see Figs.~\ref{fig:main-results-transduction}(a,b),
    independent of the spacing.  This shows that its behavior under
    different $k_0$ is similar to the single-layer case for the
    magnon-skyrmion scattering shown in Sec.~\ref{sec:SLDyn}.}
  \label{fig:transduction_hero_graphs}
\end{figure*}

Figure~\ref{fig:main-results-transduction} shows the results of a
simulation in which magnons interact across layers via a
Coulomb-mediated, interskyrmion interaction. The setup is depicted in
Figure~\ref{fig:transduction_cartoon}.  We consider two QHFMs separated
by a small insulating spacer layer, with collocated skyrmions in
each. In Figure~\ref{fig:main-results-transduction}, the skyrmions are
directly stacked one on top of the other, and held in place via an
impurity potential.  Plane magnons are driven across the skyrmion in
layer 1, which induces undulations in its core. These are transduced
via the Coulomb interaction to the skyrmion in the second layer, which
re-emits the magnons there.  The presence of a noncollinear spin
texture in both layers is necessary for Coulomb-mediated transduction,
since the Pontryagin density in Eqs.~(\ref{Pontryagin}) and
(\ref{QHMF_action}) must be nonzero in both layers.  Thus Coulomb
interactions here facilitate a novel type of ``spin drag'' effect,
with skyrmion textures functioning as the intermediary.

Each layer initially has a bare skyrmion background texture with
$Q_\text{top}=-1$ (see Appendix A). Such a profile minimizes the stiffness energy and
represents a static solution to the spin equations of motion in the
absence of Coulomb self-interaction and external effects (static in a
rotating frame in the presence of Zeeman)
\cite{girvinQuantumHallEffect1999,Rajaraman82}.  
In layer 1 (the
``transmitting layer''), magnons with wavenumber $k_0$ are driven
from the bottom row and caught by an absorbing boundary layer (ABL).
Magnons generated in layer 2 by spin drag are also dissipated by an
ABL in that layer.

In order to ensure that the topological charge distribution stays
static in the presence of Coulomb interactions and magnon scattering,
we apply an external electric pinning potential to counter the
inter-skyrmion repulsion and drift from scattering \cite{iwasakiTheoryMagnonskyrmionScattering2014}.  We incorporate an
external pinning electric field $\boldsymbol{\mathcal{E}}(x,y)$
[Eq.~(\ref{dimparams})] with a preselected (e.g.\ Gaussian)
potential and anneal the initial skyrmion textures with damping.  In some
cases, we choose an external electric field that is
exactly negative the initial internal field, and which keeps the
charge distribution in each skyrmion static, making annealing unnecessary.  For
computational efficiency, we also set the \emph{intra-layer}
interaction parameters $\alpha_1 = \alpha_2 = 0$, and retain only the
interlayer parameter $\alpha_{12} = 1$.  The effects of intralayer
Coulomb $\alpha_{1}$ and $\alpha_{2}$ are mostly negated by the
pinning potential.  The other parameters in this case are $N=140$, and
$b=0.13$, while the driven magnon wavenumber is
$k_0\approx 0.908 \, l_B^{-1}$.  A damping parameter $\damping_0=1.6$
is used to absorb the magnons at the top edge of each layer, see
Eq.~(\ref{LLG}).

\begin{figure}[b!]
  \subfigure[]{ \centering
    \includegraphics[width=0.33\textwidth]{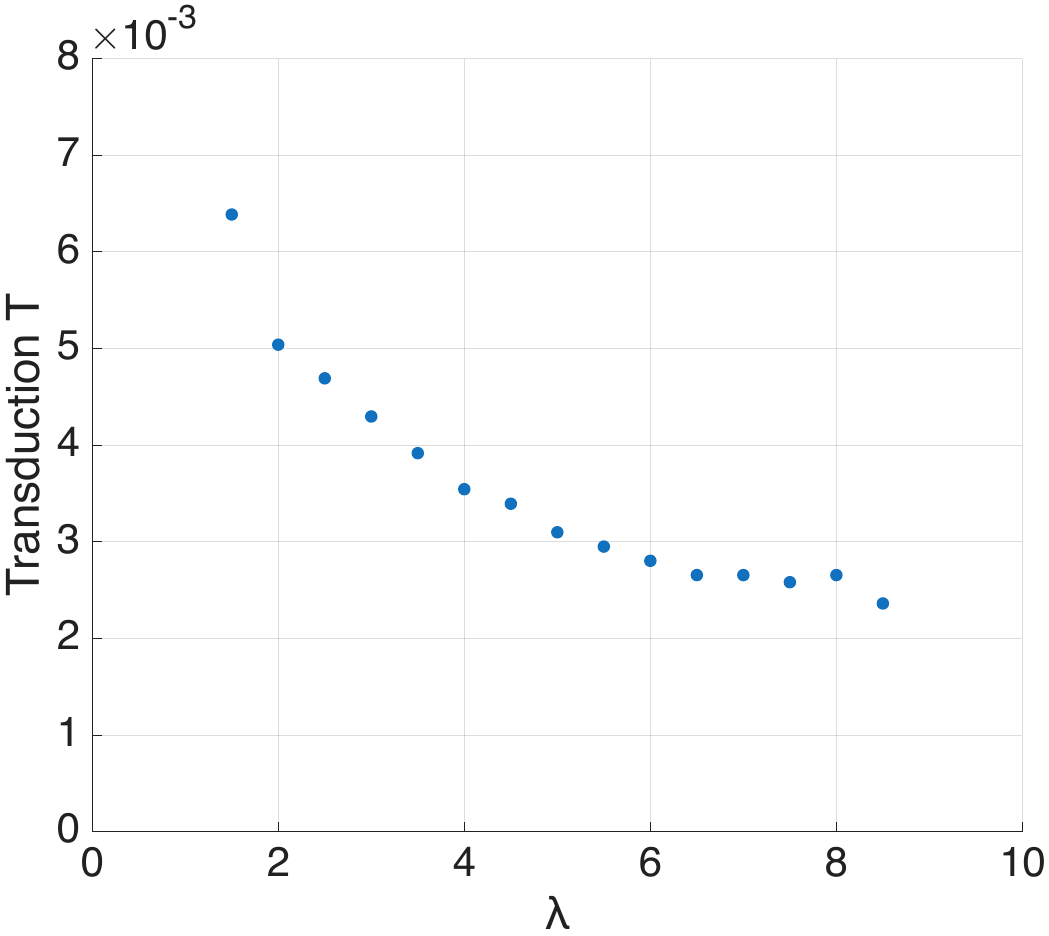}
  } \\[-1pt]
  \subfigure[]{ \centering
    \includegraphics[width=0.33\textwidth]{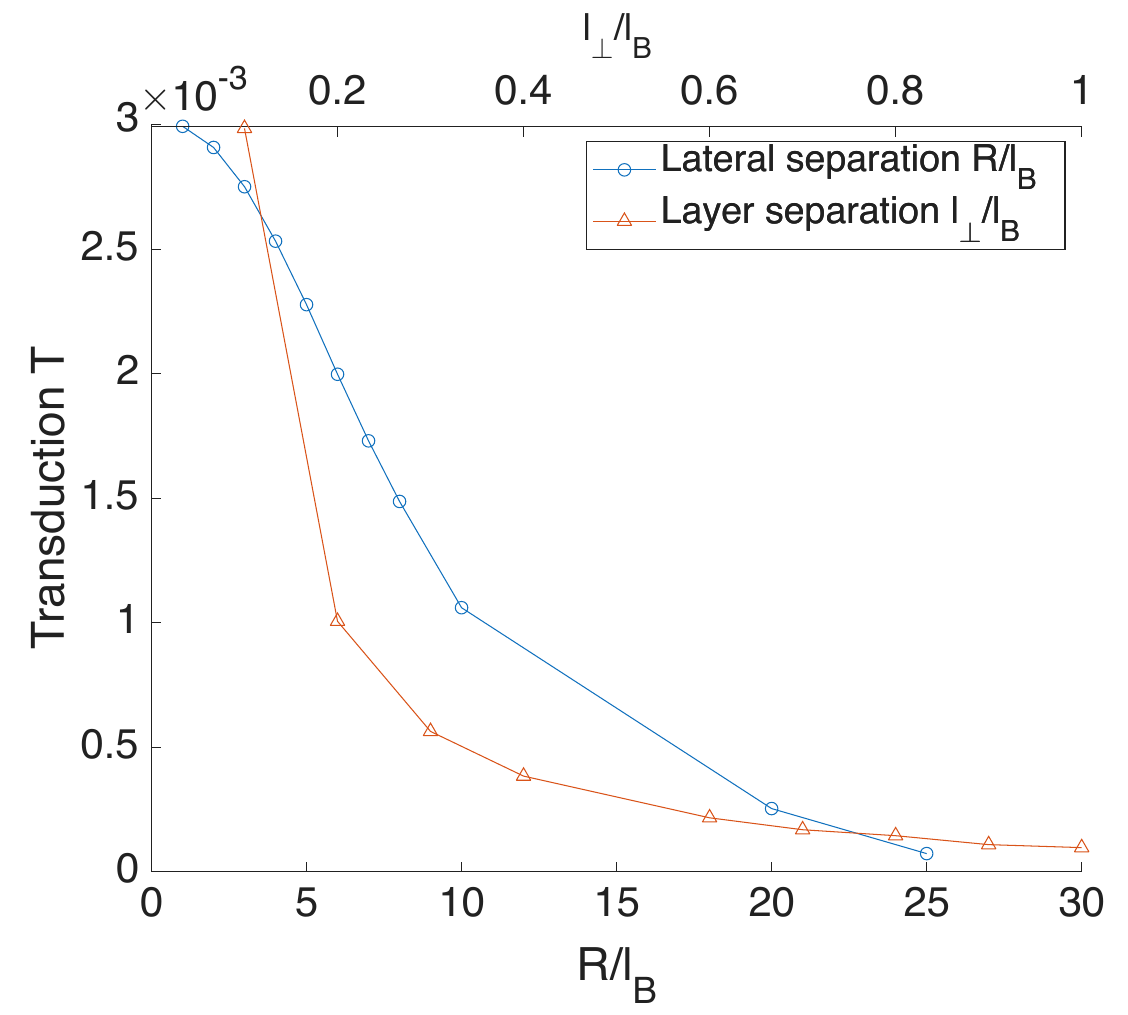}
  } \\[-1pt]
  \subfigure[]{ \centering
    \includegraphics[width=0.33\textwidth]{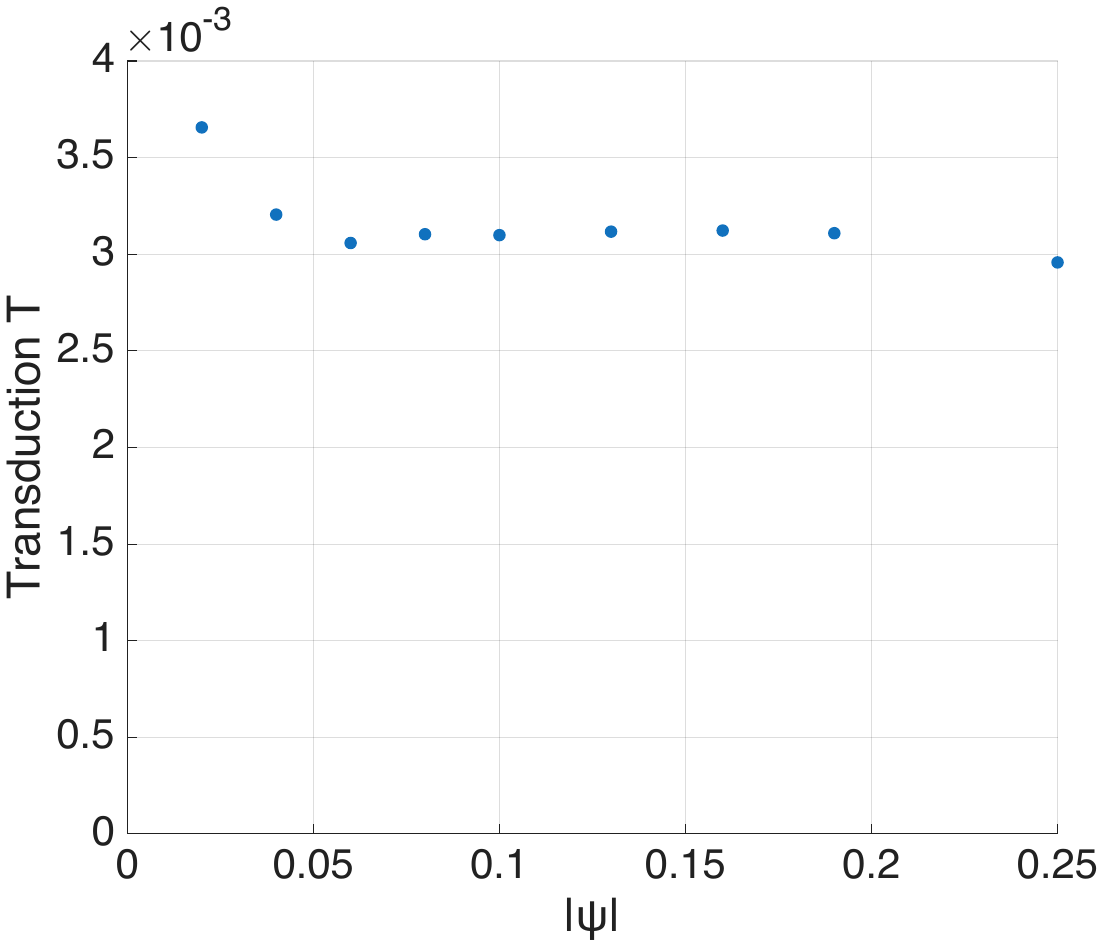}
  }
  \caption{Parameter dependence of Coulomb-mediated spin-drag
    simulations II.  (a) Transduction ratio $T$ versus skyrmion size
    $\skyrmsize$. Smaller core sizes enhance the Coulomb interaction,
    producing larger transduction ratios.  (b) Transduction ratio $T$
    versus lateral interlayer skyrmion core separation
    $R=\sqrt{\Lspace^2+l_\text{offset}^2}$ and interlayer spacing
    $\Lspace$, showing a roughly inverse relation $T\sim 1/R$ for
    large distances.  (c) Transduction ratio $T$ versus incident
    magnon amplitude $|\psi|=\sqrt{m_x^2+m_y^2}$.  Note that the ratio
    is approximately independent of perturbation magnitude, suggesting
    that transduction occurs as an effective electrodynamic linear
    response of the composite two-skyrmion, two-layer system. The
    interlayer interaction parameter $\alpha_{12}=1$ and the driven
    magnon wavenumber is $k_0 \approx 1.047 \, l_B^{-1}$.}
  \label{fig:transduction_extra_graphs}
\end{figure}
 
In layer 1, the magnons incident on the skyrmion scatter off at an
angle due to the skyrmion Hall effect \cite{garst}, as shown in
Figure~\ref{fig:main-results-transduction}(a). An otherwise free
skyrmion would also tend to drift down and to the right due to
momentum transfer from the magnon pressure, but in this case the
pinning potential holds it in place.

The novel effect of spin drag is visible in layer 2,
Figure~\ref{fig:main-results-transduction}(b).  When the driven magnons
cross over the skyrmion in layer 1, they induce magnons which emanate
from the skyrmion in layer 2.
We quantify this interlayer magnon transduction through the absorption
coefficient $\transduction$. By measuring the average power due to
damping by the ABL in each layer across a time interval near the end
of each simulation, we can compare the outgoing forward magnon flux of
each layer. $\transduction$ is the ratio of the magnon power in layer
2 to the total magnon power of both layers, see
Eq.~(\ref{transductionRT}) for a precise definition.

Our results imply that skyrmions in coupled QHFMs can act as media for
magnon transduction, or spin drag, between layers.
Figure~\ref{fig:transduction_hero_graphs}(b) shows that for a minimal
interlayer separation $\Lspace = 0.1 \, l_B$, which corresponds to an
experimentally realizable separation of $2.6$ nm \cite{Liu22spacing}
at $B = 1$ Tesla, the transduction ration $\transduction$ increases
with the driven magnon wavenumber $k_0$, showing no saturation up to
wavenumbers of order $\pi/l_B$. By contrast, larger separations
produce a decreasing $\transduction$ without a pronounced peak.  The
absence of a clear resonance is in part due to the geometry of the
detection setup that we employ here: transduced magnons are detected
only along the upper boundary of the second (receiver) layer.  As
shown in Figs.~\ref{fig:transduction_hero_graphs}(c) and
\ref{fig:k_space_difference_vs_k}, smaller wavenumbers produce more
diffuse, circular scattering in the first layer, and this is
correspondingly transduced to the second. An alternative measure of
transduced energy shown in Figure~\ref{fig:transduction_arcenergy_graph}
indicates a more pronounced decrease with increasing wavenumbers for
all but the smallest separation between layers.

It may also be that there exists a resonant frequency for spin drag
when the wavelength $\lambda=2\pi/k_0$ is comparable to the size of
the skyrmion, as is the case for the magnon Hall effect in reviewed in
Sec.~\ref{sec:SLDyn}.  For our simulations at very low $k_0$, the
scattered and transduced waves involve strong core-mode undulations of
the skyrmions that preferentially emit circular waves rather than
preserving the magnons' forward momentum in each layer.

Figure~\ref{fig:transduction_extra_graphs}(a,b) shows that the
transduction $\transduction$ is optimized when the skyrmions are
smaller and directly above each other, as the electric charge density
and core-core Coulomb energy is increased.  At the same time, spin
drag is observed at arbitrary range: the layers can be separated by
any finite distance, and the skyrmions themselves can be offset
laterally with respect to each other. The transduction coefficient
$\transduction$ appears to vary with inverse distance $1/R$ for large
$R$ in each case, as shown in
Figure~\ref{fig:transduction_extra_graphs}(b), though this point-charge
ansatz breaks down when the cores are laterally overlapping.  Finally,
Figure~\ref{fig:transduction_extra_graphs}(c) shows that the
transduction is unaffected by the strength of the layer-1 injected
magnon's incoming amplitude; roughly the same fraction of the energy
is transferred, independent of the amplitude. This suggests that the
Coulomb-mediated spin-drag phenomenon studied here arises from an
effective electrodynamic linear response in the two-layer,
two-skyrmion system.

\subsubsection{Estimation of effect size in
  experiment \label{sec:FXsize}}

To first order, the Coulomb-mediated spin drag phenomenon should scale
linearly with the number of skyrmion pairs between layers. A
particular simulation gave $T\approx0.00025$ for a pair of skyrmions
with size $\skyrmsize\approx5 \, l_B$, layer separation
$\Lspace=0.1 \, l_B$, and lateral offset $R = 20\, l_B$, with driven
magnons of wavenumber $k_0 \approx 1.047 \, l_B^{-1}$. As such, our lower estimate for $T$ in a real bilayer sample with
$\nu=1\pm0.072$ and $B=1$ Tesla is $3\times10^{-9}/\text{nm}^2$, or
$T \sim \ord{10^{-2}}$
for a sample of area $9\,\mu\text{m}^2$.

Such a phenomenon could be verified experimentally using a bilayer
version of the setup employed in
\cite{weiElectricalGenerationDetection2018}. Note that the phenomenon
of spin drag requires non-trivial topological charge in both layers,
created by doping each layer with electrons or holes. In our
simulations, if the second layer is simply in the ferromagnetic ground
state, we observe no transduction or interaction whatsoever. The spin
drag effect is maximized for the smallest separation between layers as
shown in Figure~\ref{fig:transduction_hero_graphs}(b), but the layers
should not be too close together so as to enable significant
interlayer tunneling
\cite{sternTheoryInterlayerTunneling2001,burkovPhaseTransitionSpin2002}.

\begin{figure*}[th!]
  \subfigure[]{
    \includegraphics[width=0.3\textwidth]{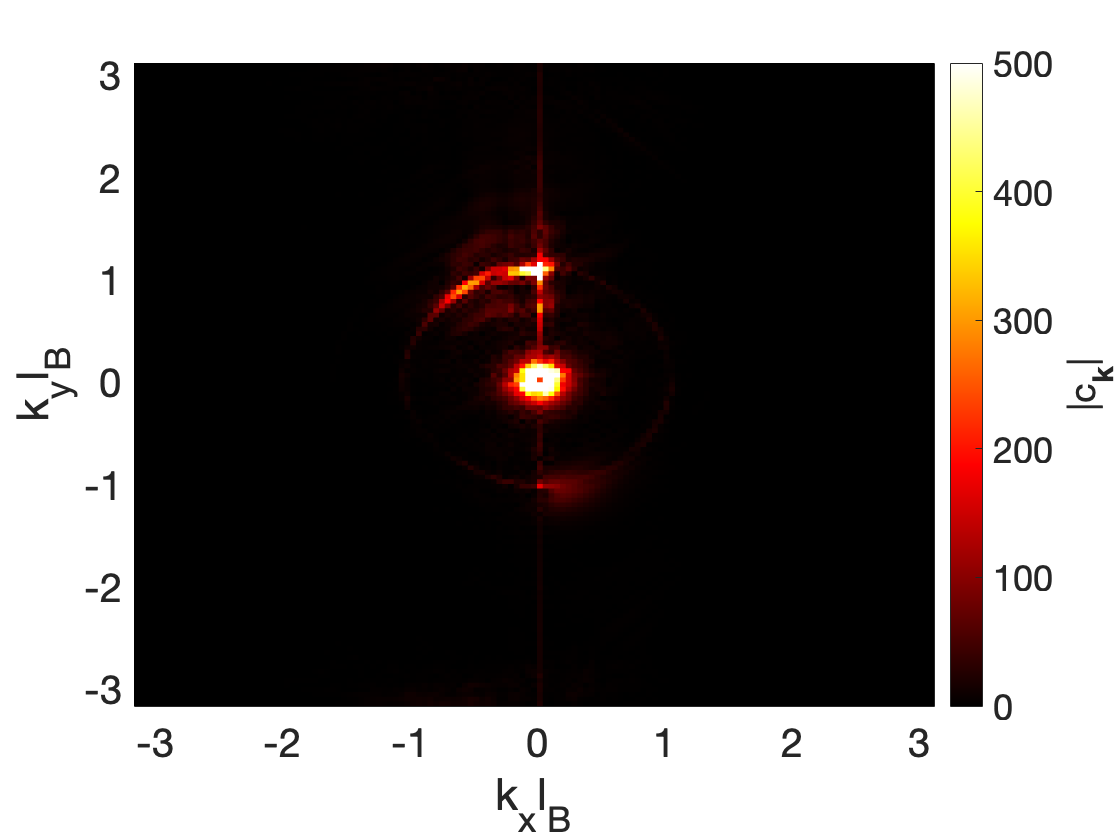}
  } \subfigure[]{
    \includegraphics[width=0.3\textwidth]{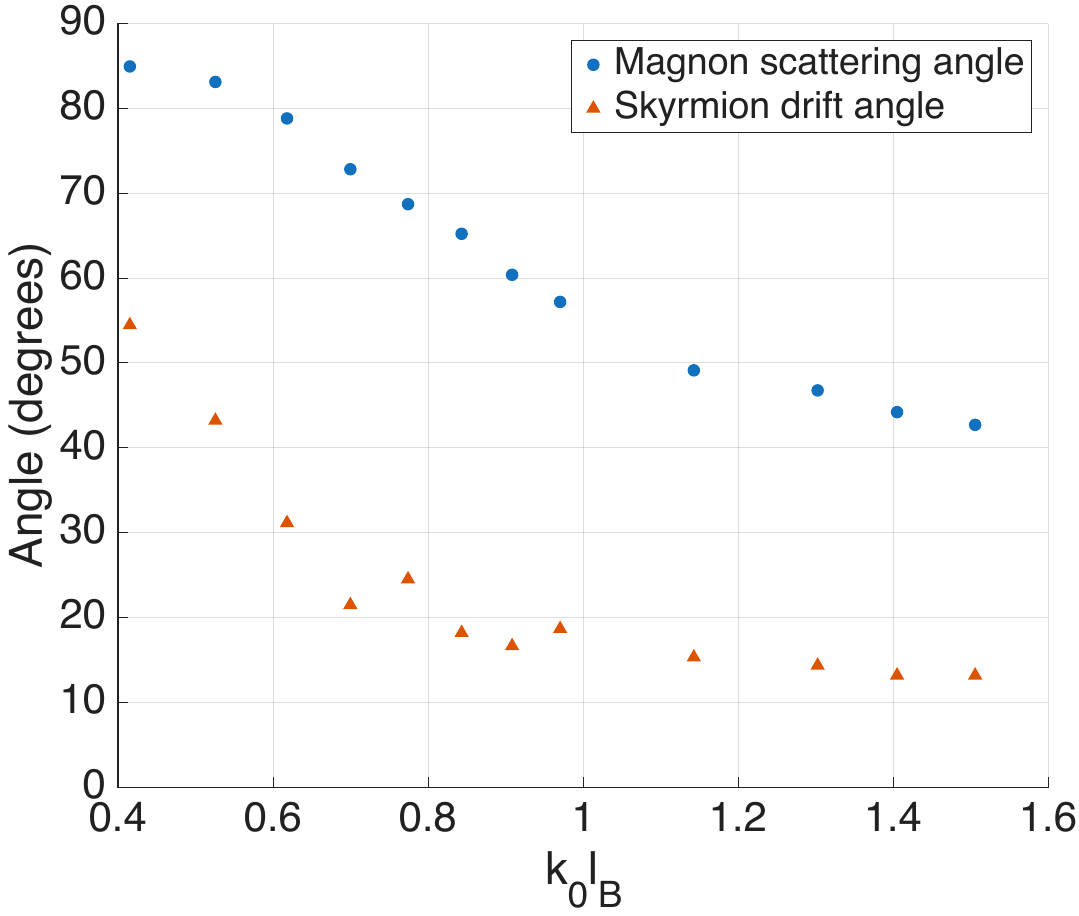}
  } \subfigure[]{
    \includegraphics[width=0.3\textwidth]{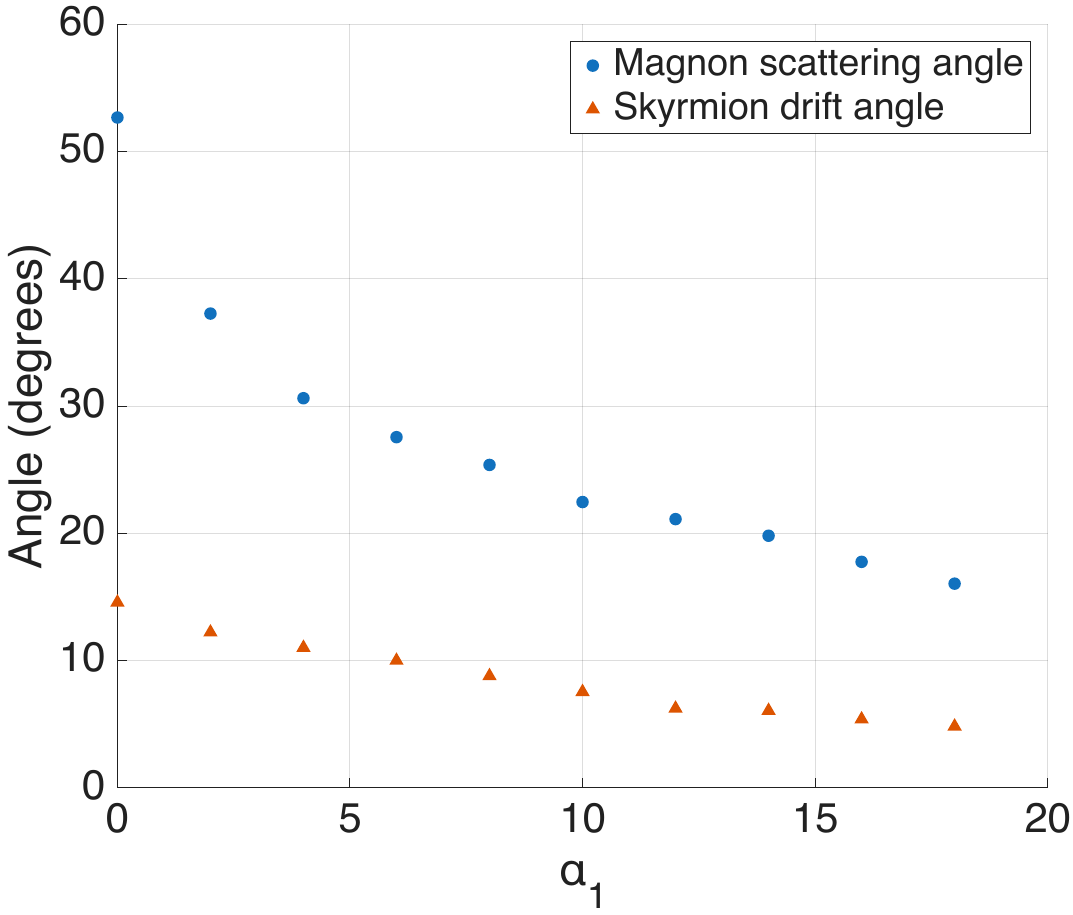}
  } \\ \hspace{1mm}
  \caption{Single-layer magnon-skyrmion scattering simulations.
    Panels (a) and (c) include Coulomb effects, while these are
    neglected in panel (b).  Results are presented for a skyrmion of
    size $\skyrmsize\approx 6 \, l_B$.  (a) $k$-space distribution of
    scattered magnons at end of a simulation, with Coulomb interaction
    parameter $\alpha=2$ and and incident magnon wavenumber
    $k_0 = 1.0472 \, l_B^{-1}$.  (b) Magnon scattering angle and skyrmion drift
    angle versus magnon driving wavevector $k_0$, with
    $\alpha=0$. Here we neglect Coulomb effects; results are presented
    to benchmark our numerics for spin-stiffness mediated interactions
    \cite{iwasakiTheoryMagnonskyrmionScattering2014,garst}.  (c)
    Magnon scattering angle and skyrmion drift angle versus Coulomb
    strength $\alpha$, with $k_0 = 1.0472 \, l_B^{-1}$. The main effect of Coulomb
    interactions is to bring both the magnon scattering angle and
    skyrmion recoil directions closer to the vertical incident line.}
  \label{fig:scattering_graphs}
\end{figure*}

\section{Numerical implementation and parameters \label{sec:MP}}

In this section, we summarize the numerical implementation of the
semiclassical spin dynamics leading to the main results presented in
Sec.~\ref{sec:MainResults}.  The setup is discussed in the context of
the bilayer spin drag calculations, although the same equations are
employed for the single-layer magnon electrodynamics
(Figure~\ref{fig:main-results-pointcharge}) and magnon-skyrmion
scattering results (Figure~\ref{fig:scattering_graphs}).

In our simulation, we adapt the equations of motion
[Eq.~(\ref{QHMF_eom})] to use numerical derivatives in position and
time:
\begin{subequations}\label{discrete_EOM}
  \begin{gather}
    \frac{\Delta\vec{m}}{\Delta\tau} =
    \vec{m}\times\vec{\mathcal{B}}_\text{eff}
    \\
    \begin{aligned}
      \mathcal{\vec{B}}_\text{eff} =&\, b\mathbf{\hat{z}}
      +\left(\vec{m}_{\mathbf{s}+\mathbf{\hat{x}}} +
        \vec{m}_{\mathbf{s}-\mathbf{\hat{x}}} +
        \vec{m}_{\mathbf{s}+\mathbf{\hat{y}}} +
        \vec{m}_{\mathbf{s}-\mathbf{\hat{y}}} \right)
      \\
      &\, -4\pi \left( \frac{\delta\varrho_{\mathbf{s}}}{\delta
          \vec{m}}V_\mathbf{s} +
        \frac{\delta\varrho_{\mathbf{s}-\mathbf{\hat{x}}}}{\delta
          \vec{m}}V_{\mathbf{s}-\mathbf{\hat{x}}} \right)
      \\
      &\, -4\pi \left(
        \frac{\delta\varrho_{\mathbf{s}-\mathbf{\hat{y}}}}{\delta
          \vec{m}}V_{\mathbf{s}-\mathbf{\hat{y}}} +
        \frac{\delta\varrho_{\mathbf{s}-\mathbf{\hat{x}}-\mathbf{\hat{y}}}}{\delta
          \vec{m}}V_{\mathbf{s}-\mathbf{\hat{x}}-\mathbf{\hat{y}}}
      \right)
    \end{aligned}
  \end{gather}
\end{subequations}
Here we employ the dimensionless parameterization introduced in
Eq.~(\ref{dimparams}); the lattice spacing in position is the magnetic
length $l_B$.

The continuous magnetization field in position space
$\vec{m}(\vex{r})$ is discretized over a square lattice, with lattice
vector $\mathbf{r} = l_B \, \mathbf{s}$.  Here $[s_x,s_y]$ denote the
integer coordinates of a lattice site. In Eq.~(\ref{discrete_EOM}),
\begin{align}
  \varrho_\mathbf{s} \equiv l_B^2 \, \varrho[\vec{m}(\tau,\mathbf{r}= l_B \, \mathbf{s})]
\end{align}
is the discretized Pontryagin density, not to be confused with the
stiffness constant $\rho_s$ used to set the energy units
[Eq.~(\ref{dimparams})].  By replacing the triple product with a solid
angle calculation on the unit sphere
\cite{vanoosteromSolidAnglePlane1983}, we can find the charge density
exactly in the discretized system.  We choose a gauge where the scalar
potential at a plaquette to the top-right of site $\mathbf{s}$ is
\begin{align}
  V_\mathbf{s}=-\boldsymbol{\mathcal{E}}\cdot\mathbf{s} + \alpha \sum\limits_{\mathbf{s'\neq s}}\frac{\varrho_{\mathbf{s'}}}{|\mathbf{s-s'}|} \label{}
\end{align}
and sum the electric contributions due to all four plaquettes
neighboring the site.  Time evolution is carried out via the
fourth-order Runge-Kutta method.

All magnons in our discretized code obey the dispersion:
\begin{align}
  \Omega(\mathbf{k}) = b + 4 - 2\cos(k_x l_B) - 2\cos(k_y l_B),
  \label{dispersion}
\end{align} 
where frequency $\Omega$ has units of inverse dimensionless time
$\tau$.  This only approximates the continuum quadratic
dispersion reasonably for low $\mathbf{k}$, so we only use low-energy magnons
in the simulations.

Note that as this solution is energy conserving, the spins follow a
path along an equipotential in the presence of the effective magnetic
field rather than canting to align with it. Similarly, the charge
tends to move perpendicularly to the effective electric field at
constant velocity rather than along it. To locally minimize the
energy, we can add a dissipative term to the equations of motion:
\begin{align}
  \frac{d\vec{m}}{d\tau}
  &=
    \vec{m}\times\vec{\mathcal{B}}_\text{eff} 
    - 
    \damping \,
    \vec{m}\times(\vec{m}\times\vec{\mathcal{B}}_\text{eff}), 
    \label{LLG}
\end{align}
where $\damping$ is the damping strength. This form can be derived
from the Landau-Lifshitz-Gilbert equation, which has been shown to
phenomenologically describe dissipative effects in
micromagnetics \cite{ahoroni}.

The system's change in total energy over time can be found with
\begin{align}
  \frac{dE}{dt}&=\frac{d\vec{m}}{dt}\cdot\vec{\mathcal{B}}_\text{eff}. 
                 \label{totalEchange}
\end{align}
Using the LLG equation \eqref{LLG}, we can find the change in energy
specifically due to dissipative effects:
\begin{align}
  \frac{dE}{d\tau}&=
                    -
                    \damping
                    \left[
                    (\vec{\mathcal{B}}_\text{eff}\cdot\vec{m})^2-|\vec{\mathcal{B}}_\text{eff}|^2
                    \right]
\end{align}
From this, by measuring the power of dissipation due to the
absorbing boundary layers (ABLs)
of each layer separately, we can compare the intensities of magnons
scattered from the skyrmions into each layer, specifically from the
driving layer (layer 1) to the receiving layer (layer 2).  We define
the transduction ($\transduction$) and transmission ($\transmission$)
coefficients as follows,
\begin{subequations}\label{transductionRT}
  \begin{align}
    \transduction
    &=
      \left(\frac{d E_\text{L2}}{d\tau}\right)/\left(\frac{dE}{d\tau}\right),
    \\
    \transmission
    &=
      1-\transduction,
  \end{align}
\end{subequations}
where $E_\text{L2}$ denotes the energy absorbed in layer 2, while $E$
is the total energy absorbed in both layers.  However, since the
skyrmion texture is not completely localized, the ABL in each layer
can absorb energy from the tails of the skyrmion configurations, even
if the texture is first subject to annealing (damping everywhere).  As
such, to isolate the power due to magnons alone in each simulation, we
subtract the power absorbed from an identical simulation
\emph{without} driven spins.

\section{Single-layer dynamics \label{sec:SLDyn}}

In this section we describe the analytical approach to magnon-charge
scattering. We also present additional numerical results for
magnon-skyrmion scattering in both the presence and absence of Coulomb
interactions. The latter are used to benchmark our numerics against
previous studies.

\subsection{Magnon-point charge interactions}

We can expand Eq.~(\ref{QHMF_action}) to second order in magnon
fluctuations around a ferromagnetic ground state, leading to the
action
\begin{subequations}
  \begin{align}
    S_\Pi 
    &\simeq 
      \int d t \, d^2\vex{r} 
      \left\{
      \bar{\Pi}\left[i s n \pd_t + \rho_s\nabla^2 + s n g \mu B\right]\Pi
      \right\} 
    \\
    &+
      \int d t \, d^2\vex{r} \,     
      \vec{E}\cdot\left(\frac{e}{4\pi}\vec{J}\times\hat{z}\right),
      \label{MagnonElecEq}
  \end{align}
\end{subequations}
where $\Pi\equiv \frac{1}{\sqrt{2}}\left(m^x + i m^y\right)$ is the
complex boson field operator for magnons, and
\begin{align}\label{spincurrent}
  \vec{J}
  \equiv 
  -\frac{i}{2}\bar{\Pi}\left(\overrightarrow{\nabla}-\overleftarrow{\nabla}\right)\Pi
\end{align}
is the Noether current associated to U(1) rotational invariance around
the $z$-axis.  Note that as the electrical potential term now
manifests as a coupling between the \emph{electric field} and the
\emph{magnon current} (both time-reversal even, polar-vector
quantities), rather than the topological charge.  Magnons carry an
effective electric dipole moment density
$\vec{d}\equiv\frac{e}{4\pi}\vec{J}\times\hat{z}$ and can scatter off
of non-uniform electric fields despite possessing zero net Pontryagin
(electric) charge.

Rescaled in terms of our dimensionless simulation parameters and with
lattice spacing $l_B=1$, an effective single-particle Hamiltonian can
be extracted from the action:
\begin{subequations}
  \begin{align}
    S_\Pi &= \frac{1}{4\pi}\int\limits_{\tau,\mathbf{r}}\bar{\Pi}\left(i\pd_{\tau}-\hat{h}\right)\Pi \\
    \hat{h} &= b - \nabla^2 - \frac{i\e^{\beta\alpha}}{2}\left[\pd_\beta\mathcal{E}^\alpha(\hat{\mathbf{r}}) +\mathcal{E}^\alpha(\hat{\mathbf{r}}) \pd_\beta\right] \\
          &= b + \frac{1}{2\mathcal{M}}\left[\mathbf{\hat{P}}-\boldsymbol{\mathcal{A}}(\hat{\mathbf{r}})\right]^2 + \mathcal{V}(\mathbf{\hat{r}}), 
            \label{MagnonHamiltonianSingle}
  \end{align}
\end{subequations}
where $\mathcal{M}=1/2$ is the magnon mass,
$\hat{\mathbf{P}}=-i\nabla$ is the magnon momentum operator, and
$\mathcal{\boldsymbol{A}}(\mathbf{r})$, $\mathcal{V}(\mathbf{r})$ are
the synthetic vector and scalar potentials for the magnon due to the
physical electric field $\vex{\mathcal{E}}(\vex{r})$, defined via
Eq.~(\ref{SyntheticPotentials}).

To find the wave function of a magnon in position space subject to the
perturbing Hamiltonian
$ \hat{h}_1\equiv
-\left[\mathbf{\hat{P}}\cdot\boldsymbol{\mathcal{A}}(\hat{\mathbf{r}})
  + \boldsymbol{\mathcal{A}}(\hat{\mathbf{r}})\cdot
  \mathbf{\hat{P}}\right] $, we must solve the Schr\"odinger equation:
\begin{align}
  \left\{(\nabla^2+\omega) -i\left[\nabla\cdot\boldsymbol{\mathcal{A}}(\hat{\mathbf{r}}) + \boldsymbol{\mathcal{A}}(\hat{\mathbf{r}})\cdot \nabla\right]\right\}\psi=0,
\end{align}
where $\omega=k^2+i\eta$ is the on-shell frequency of an unperturbed
magnon, and the energy shift $b$ is ignored. A formal solution is
\begin{align}\label{ScattSol}
  \psi(\mathbf{r}) 
  &= 
    \psi_0(\mathbf{r}) 
    + 
    2 i 
    \int
    d^2\vex{r'}
    \,
    G(\omega,\mathbf{r-r'})
    \left[\boldsymbol{\mathcal{A}}(\mathbf{r'})\cdot\nabla'\right]
    \psi(\mathbf{r'}),
\end{align}
where $\psi_0(\vex{r})$ denotes a solution to the unperturbed problem,
and
\begin{align}
  G(\omega,\vex{r})
  \equiv
  \frac{1}{4i}H_0^{(1)}(r\sqrt{\omega})
\end{align}
is the retarded free-space Green's function for 2+1-D ferromagnetic
outgoing magnons. In the above, $H_0^{(1)}$ is the Hankel function of
the first kind, and we have used the fact that
$\nabla\cdot\boldsymbol{\mathcal{A}}=0$ in this case.

\begin{figure*}[th!]
  \subfigure[]{ \centering
    \includegraphics[width=0.31\textwidth]{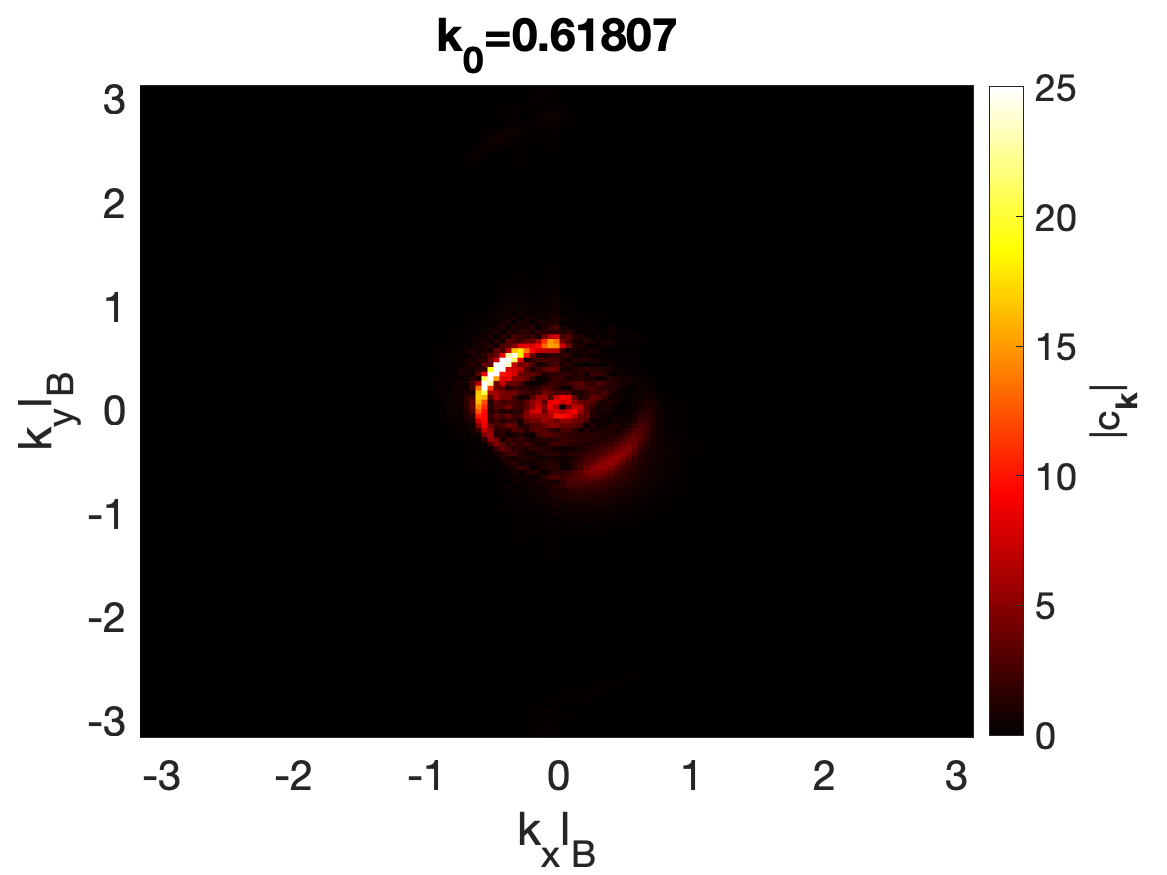}
  } \subfigure[]{ \centering
    \includegraphics[width=0.31\textwidth]{Images/FFT_L2diff_final0.9.png}
  } \subfigure[]{
    \includegraphics[width=0.31\textwidth]{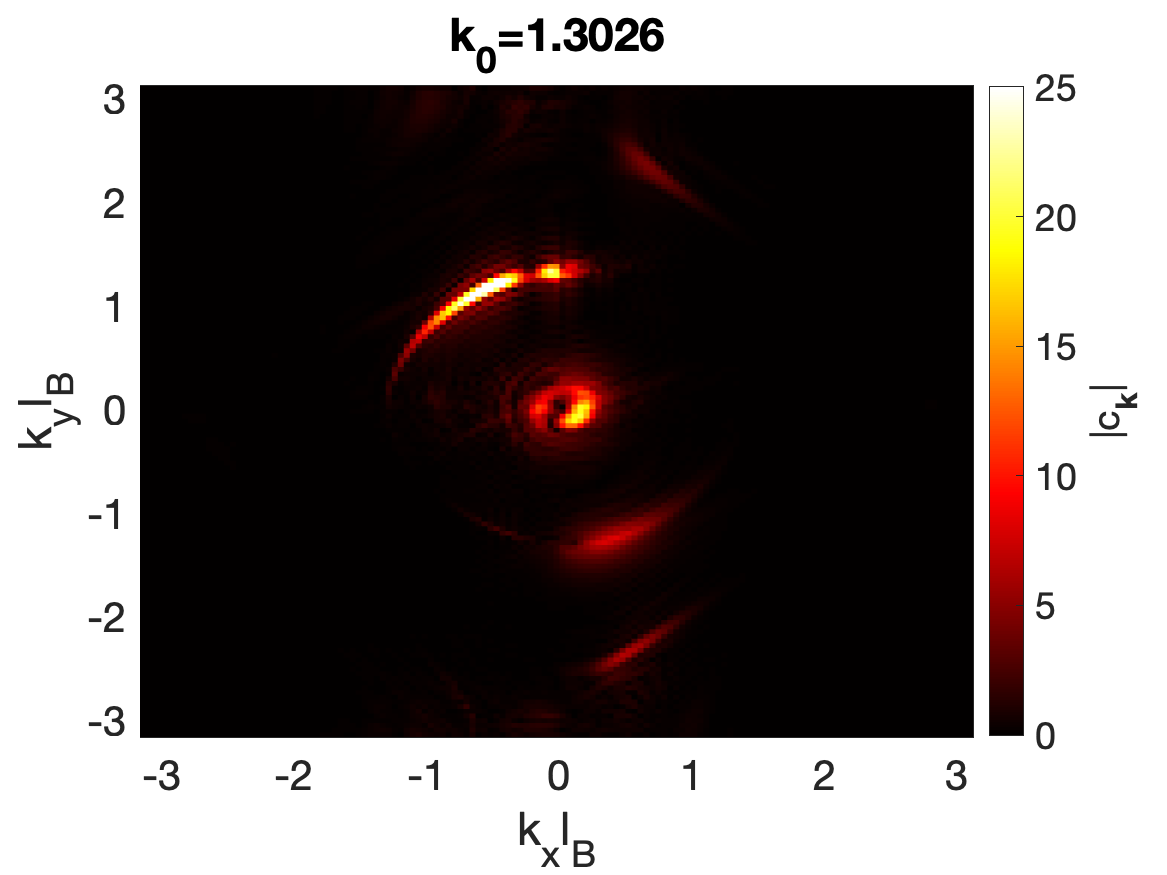}
  } \hspace{1mm}
  \caption{Coulomb-mediated spin-drag simulations III.  The plots
    depict the $k$-space distribution of magnons in the receiving
    layer at the end of a driven simulation, with skyrmion size $\skyrmsize=5\,l_B$, interaction parameter $\alpha=1$, interlayer separation $\Lspace=0.1 \, l_B$,
    and Gaussian pinning potential in each layer. As in
    Figure~\ref{fig:transduction_hero_graphs}(a), the results of the
    undriven second layer (with stable skyrmion texture) are
    subtracted here to highlight the spin drag. Results are depicted
    for three different layer-one driven magnon wave vectors $k_0$.
    These panels show that the emitted magnons give a more diffuse,
    circular scattering pattern at smaller wave numbers. This is
    partly responsible for the absence of a clear resonance in the
    transduction versus $k_0$ plot,
    Figure~\ref{fig:transduction_hero_graphs}(b).}
  \label{fig:k_space_difference_vs_k}
\end{figure*}

The first Born approximation for the scattered wave amplitude
$\delta\psi_B^{(1)}(\mathbf{r})$ obtains by replacing $\psi(\vex{r'})$
on the right-hand side (RHS) of Eq.~(\ref{ScattSol}) by the
free-propagating wave $\psi_0(\vex{r'})$, while the second Born
amplitude $\delta\psi_B^{(2)}$ replaces
$\psi \rightarrow \delta\psi_B^{(1)}$ on the RHS of
Eq.~(\ref{ScattSol}). Together, these give
\bsub\label{SecondBornEq}
\begin{align}
  \psi(\mathbf{r}) 
  \simeq&\,
          \psi_0(\mathbf{r}) 
          + 
          \delta\psi_B^{(1)}(\mathbf{r}) 
          + 
          \delta\psi_B^{(2)}(\mathbf{r}) 
          + 
          \ord{k\lambdaE}^3, 
\end{align}
where the on-shell amplitude is
\begin{align}
  \psi_0(\mathbf{r})
  =&\,
     e^{ikr\sin\phi},
  \\
  \label{FirstBorn}
  \delta\psi_B^{(1)}(\mathbf{r}) 
  =&\, 
     -
     \frac{\pi k\lambdaE}{2\sqrt{2}}
     \frac{\cos\phi}{\sqrt{1-\sin\phi}}H_0^{(1)}(kr), 
  \\
  \delta\psi_B^{(2)}(\mathbf{r}) 
  =&\, 
     -
     \frac{\pi(k\lambdaE)^2}{16}
     f(\phi)
     \,
     H_0^{(1)}(kr)
\end{align}
\begin{align}
  f(\phi)
  =&\,
     \int_0^{2 \pi} d\phi'
     \sqrt{
     \left[1-\cos(\phi-\phi')\right]^3
     \left[1-\sin(\phi')\right]
     }. 
\end{align}
\esub Here we assume an incident plane wave state
$\psi_0 = e^{i k y}$.  We note that the scattering from a charge in
the plane of the magnon produces a non-analytic form factor already in
the first-Born approximation, Eq.~(\ref{FirstBorn}).  The magnitude of
the wavefunction in Eq.~(\ref{SecondBornEq}) is visualized in
Figure~\ref{fig:main-results-pointcharge}(b), for $k=1.5$ and
$\lambdaE=-0.4$.

This expansion is valid for $k\lambdaE\ll1$, and in this realm the
model clearly resembles our numerical findings.  The first-order term
in $k\lambdaE$ modifies the plane wave such that the wave-function
magnitude varies across space in a distribution similar to
Figure~\ref{fig:main-results-pointcharge}(a).  The second-order
correction $\delta\psi_B^{(2)}(\mathbf{r})$ (significant for strong
point charges) magnifies the outgoing Hankel function to one side,
corresponding with the sign of the charge.

\subsection{Magnon-skyrmion interactions}

In chiral magnets, magnons are known to scatter when passing over a
skyrmion background texture, and to impart a recoil drift velocity to
the impacted skyrmion
\cite{iwasakiTheoryMagnonskyrmionScattering2014}.  In this subsection,
we specialize our code to the single-layer case to demonstrate that
this phenomenon also exists in QHFMs.

\subsubsection{Magnon and skyrmion Hall angle vs $k$}

It has been shown numerically
\cite{iwasakiTheoryMagnonskyrmionScattering2014} that in a ferromagnet
with Dzyaloshinskii-Moriya interactions, the magnon scattering angle
and skyrmion drift angle are maximized when the incident magnon
wavelength equals the skyrmion size. In addition, due to conservation
of momentum, the magnon scattering angle should always be twice the
skyrmion drift angle. In our case, it is difficult to judge the
deflection angle at this resonant frequency, as the incident magnons
excite internal modes within the skyrmion that emit circular
waves. However, the skyrmion drift angle is maximized at this
frequency, and the conservation ratio roughly holds for higher magnon
wavenumbers, as shown in Figure~\ref{fig:scattering_graphs}(b).

\subsubsection{Magnon and skyrmion Hall angle vs $\alpha$}

We also vary the Coulomb strength $\alpha$ in these single-layer
scattering simulations to discern the influence of this interaction
unique to QHFMs.  The skyrmion is first left to expand and
equilibrate, and then magnons are driven.  We find that increasing the
Coulomb interaction parameter $\alpha$ slightly decreases both the
scattering angle and the skyrmion drift angle, as shown in
Figure~\ref{fig:scattering_graphs}(c). However, large values of $\alpha$
produce unreliable results, as they allow for magnon-magnon
interactions that produce a variety of different wavenumbers. The
angle variation could be influenced by the skyrmion's size varying
with $\alpha$, but earlier results from Sec.~\ref{sec:MainResults}
show that magnons interact with the electric field a skyrmion produces
as well as its texture, which is non-negligible here.

\section{Conclusion \label{sec:Conc}}

Using semiclassical simulations, we have examined two effects that
arise from Coulomb interactions in the magnetization dynamics of
QHFMs.  First, we demonstrated magnon deflection by an electric
charge, Figure~\ref{fig:main-results-pointcharge}. This arises due to
the magnon electric dipole moment, which is proportional but
perpendicular to the spin current.

Second, we observe a Coulomb-mediated ``spin drag'' effect between
disconnected, adjacent layers, Figs.~\ref{fig:transduction_cartoon}
and
\ref{fig:main-results-transduction}--\ref{fig:transduction_extra_graphs}.
The effect occurs due to Coulomb-coupling between undulations in the
topological textures of both layers, induced by magnon injection in
one layer and observed as magnon generation in the other.  The degree
of power transduction between layers was quantified in terms of skyrmion
densities achievable by small doping away from $\nu = 1$, see
Sec.~\ref{sec:FXsize}.

Multiple avenues exist for further studies.  First, these ideas can be
generalized to other types of quantum Hall magnets with higher
symmetries and more complicated defects, e.g.\ SU($N$) skyrmions
\cite{Arovas99}.  Second, one can also explore the dynamics of
magnetic textures in QHFM analogs reported in moir\'e materials
\cite{Cao18ins,Sharpe19,Lu20}.  Another direction is to incorporate
quantum fluctuations, which have been neglected in this work.

\acknowledgments

We thank Yonglong Xie for helpful discussions that partly inspired
this work, and Yunxiang Liao for early versions of the numerical
code. This work was supported by the Welch Foundation Grant No.~C-1809
(A.C. and M.S.F.) This work was supported in part by the Big- Data
Private-Cloud Research Cyberinfrastructure MRI-award funded by NSF
under Grant No.~CNS-1338099 and by Rice University’s Center for
Research Computing (CRC).

\appendix

\section{Skyrmion texture parameterization\label{app:skyrmion}}

Retaining only the stiffness interaction [i.e., setting the Coulomb
interaction parameter $\alpha = 0$, Eq.~(\ref{dimparams})], the
magnetization field for a stable ``bare'' skyrmion with $Q_\text{top}=-1$ can
be written as \cite{Rajaraman82,belavinMetastableStatesTwodimensional1975}
\begin{align}\label{skyrmionsize}
  \vec{m}(r,\phi)=\left[\frac{4\skyrmsize r\cos\phi}{r^2+4\skyrmsize^2},\frac{4\skyrmsize r\sin\phi}{r^2+4\skyrmsize^2},\frac{r^2-4\skyrmsize^2}{r^2+4\skyrmsize^2}\right],
\end{align}
where ($r,\phi$) are polar coordinates in the sample plane, the core
is centered at the origin, and $\skyrmsize$ denotes half of the
skyrmion radius. The inclusion of other interactions deforms
the bare stable texture \cite{Fertig94}; in our spin
simulations, the Coulomb interaction makes an initially bare
skyrmion profile expand a bit before stabilizing.

\begin{figure}
  \subfigure[]{
    \includegraphics[width=0.35\textwidth]{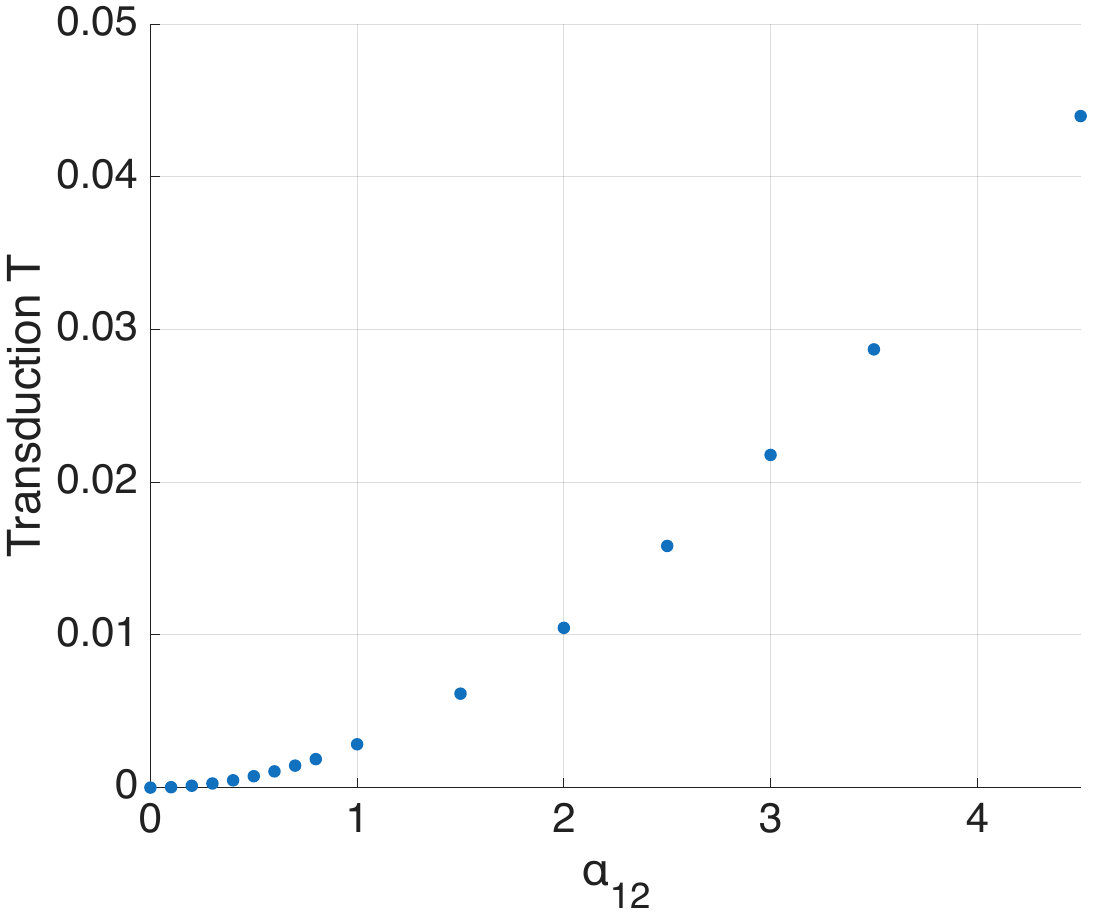}
  } \\ \hspace{1mm}
  \caption{Transduction ratio $T$ vs varied interlayer Coulomb
    strength $\alpha_{12}$, with skyrmion size $\skyrmsize=5\,l_B$, constant
    Gaussian pinning potential in each layer, driven magnon wavenumber
    $k_0 \approx 1.047 \, l_B^{-1}$, and interlayer spacing
    $\Lspace=0.1 \, l_B$.}
  \label{fig:transduction_alpha_graph}
\end{figure}

\begin{figure}
  \subfigure[]{
    \includegraphics[width=0.35\textwidth]{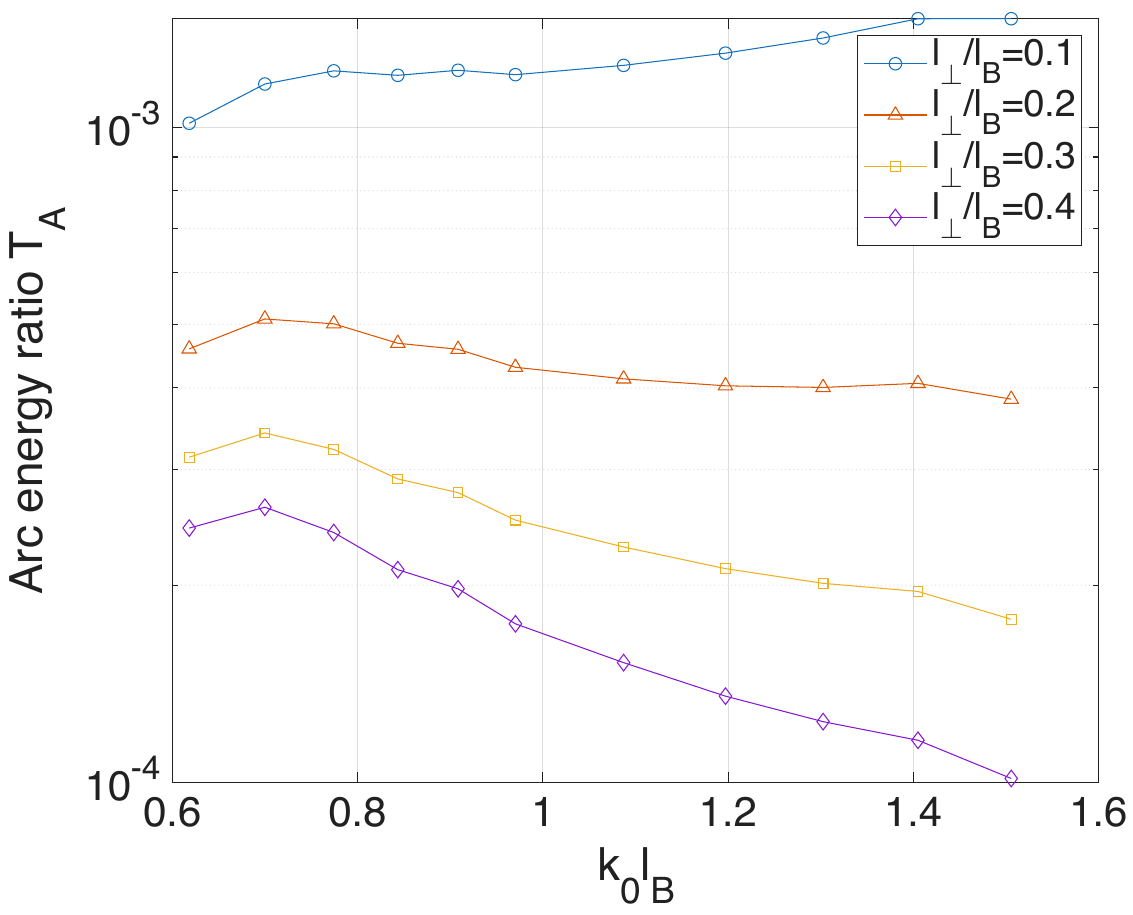}
  } \\ \hspace{1mm}
  \caption{Transduction ratio $T_A$ (captured by arc-energy ratio, see
    Appendix~\ref{app:SM}) versus driven magnon wavenumber $k_0$ and
    interlayer spacing $\Lspace$ for various simulations, with
    skyrmion size $\skyrmsize=5\,l_B$, Coulomb strength $\alpha=1$, and
    Gaussian pinning potential in each layer.}
  \label{fig:transduction_arcenergy_graph}
\end{figure}

\section{Supplemental graphs\label{app:SM}}

Figure~\ref{fig:k_space_difference_vs_k} shows the magnitude of the $k$-space distribution of magnons in the
second (``receiving'') layer at the end of a transduction simulation, for varied driving wavenumbers
$k_0$ [see Figure~\ref{fig:transduction_hero_graphs}(b)]. The result of a ``base case'' without driven magnons is subtracted to graphically reduce the strong standing modes of
the skyrmion, so that the weak transduced magnon modes are relatively
brighter.  The modes seen near the origin are due to the phase shift
from the skyrmion drifting and deforming.  We can see that the
higher-momentum driven magnons glance off of the skyrmion at a smaller
angle, but with a similar profile.

Figure~\ref{fig:transduction_alpha_graph} shows the transduction ratio
$T$ resulting from simulations with varied interlayer Coulomb strength
$\alpha_{12}$. The graph suggests a nonlinear relationship, perhaps
that the interlayer magnon interaction term is second-order in
$\alpha$. In reality, this parameter is a fixed number of
order one, because the spin stiffness is itself mediated by Coulomb
exchange.

Figure~\ref{fig:transduction_arcenergy_graph} depicts an alternative
means of quantifying interlayer magnon transduction, compare to the
fixed detector geometry assumed in
Figure~\ref{fig:transduction_hero_graphs}(b).  Here, the transduction
ratio $T_A$ is calculated in a different manner so as to account for
all outgoing magnon modes equally.  Whereas $T$
[Figure~\ref{fig:transduction_hero_graphs}(b)] is the fraction of the
power caught by the ABL in layer 2, favoring forward momentum as an
experimental setup might, $T_A$
[Figure~\ref{fig:transduction_arcenergy_graph}] is the fraction of the
energy in layer 2 obtained by summing the energy contributions of
every mode on the $|\mathbf{k}|=k_0$ arc in $k$-space within
$-\pi/4<\phi<3\pi/4$, each contribution of which is proportional to
its amplitude squared: $E(\mathbf{k})\sim|c_\mathbf{k}|^2$.  This is
intended to more accurately describe low-$k_0$ cases where the
scattering profile is mostly circular to the left
[Figure~\ref{fig:k_space_difference_vs_k}], but the graph shows a mostly
similar result: a positive $T_A$ vs $k_0$ correlation for low
interlayer spacing and a negative one for most others.


\begin{thebibliography}{99}
\bibitem{girvinQuantumHallEffect1999} S. M. Girvin, The Quantum Hall
  Effect: Novel Excitations and Broken Symmetries, in
  \emph{Topological Aspects of Low Dimensional Systems}, edited by
  A. Comtet, T. Jolicoeur, S. Ouvry, and F. David (Springer, Berlin,
  Germany, 1999);
  \href{https://doi.org/10.48550/arXiv.cond-mat/9907002}{arXiv:cond-mat/9907002}.
\bibitem{Sondhi93} S. L. Sondhi, A. Karlhede, S. A. Kivelson, and
  E. H. Rezayi, Skyrmions and the crossover from the integer to
  fractional quantum Hall effect at small Zeeman energies,
  \href{https://doi.org/10.1103/PhysRevB.47.16419}{Phys. Rev. B {\bf
      47}, 16419 (1993).}
\bibitem{Fertig94} H. A. Fertig, L. Brey, R. C\^ot\'e, and
  A. H. MacDonald, Charged spin-texture excitations and the
  Hartree-Fock approximation in the quantum Hall effect,
  \href{https://doi.org/10.1103/PhysRevB.50.11018}{Phys. Rev. B {\bf
      50}, 11018 (1994).}
\bibitem{Fertig97} H. A. Fertig, L. Brey, R. C\^ot\'e,
  A. H. MacDonald, A. Karlhede, and S. L. Sondhi, Hartree-Fock theory
  of skyrmions in quantum Hall ferromagnets,
  \href{https://doi.org/10.1103/PhysRevB.55.10671}{Phys. Rev. B {\bf
      55}, 10671 (1997).}
\bibitem{bychkovTWODIMENSIONALELECTRONSSTRONG1981} Yu. A. Bychkov, S. V. Iordanskii and G. M. Eliashberg, Two-dimensional electrons in a strong magnetic field,
  Pis’ma Zh. Eksp. Teor. Fiz {\bf 33}, 152 (1981) [JETP Lett. {\bf
    33}, 143 (1981)].
\bibitem{Kallin84} C. Kallin and B. I. Halperin, Excitations from a filled Landau level in the two-dimensional electron gas,
  \href{https://doi.org/10.1103/PhysRevB.30.5655}{Phys. Rev. B {\bf
      30}, 5655 (1984).}
\bibitem{gorkovContributionTheoryMott1968} L. P. Gor'kov and I. E. Dzyaloshinskii, Contribution to the Theory of the Mott Exciton in a Strong Magnetic Field,
  Zh. Eksp. Teor. Fiz. {\bf 53}, 717 (1967) [Sov. Phys. JETP {\bf
    26}, 449 (1968)].
\bibitem{Lerner78} I. V. Lerner and Yu. E. Lozovik, Mott exciton in a
  quasi-two-dimensional semiconductor in a strong magnetic field,
  Zh. Eksp. Teor. Fiz. {\bf 78}, 1167 (1978) [Sov. Phys. JETP {\bf
    51}, 588 (1980)].
\bibitem{Cote97} R. C\^ot\'e, A. H. MacDonald, L. Brey, H. A. Fertig,
  S. M. Girvin, and H. T. C. Stoof, Collective Excitations, NMR, and
  Phase Transitions in Skyrme Crystals,
  \href{https://doi.org/10.1103/PhysRevLett.78.4825}{Phys. Rev. Lett. {\bf
      78}, 4825 (1997).}
\bibitem{Sinova00} J. Sinova, A. H. MacDonald, and S. M. Girvin,
  Disorder and interactions in quantum Hall ferromagnets near
  $\nu = 1$,
  \href{https://doi.org/10.1103/PhysRevB.62.13579}{Phys. Rev. B {\bf
      62}, 13579 (2000).}
\bibitem{barrettOpticallyPumpedNMR1995} S. E. Barrett, G. Dabbagh,
  L. N. Pfeiffer, K. W. West and R. Tycko, Optically Pumped NMR
  Evidence for Finite-Size Skyrmions in GaAs Quantum Wells near
  Landau Level Filling $\nu = 1$,
  \href{https://doi.org/10.1103/PhysRevLett.74.5112}{Phys. Rev. B {\bf
      74}, 5112 (1995).}
\bibitem{Yang06} K. Yang, S. Das Sarma, and A. H. MacDonald,
  Collective modes and skyrmion excitations in graphene SU(4) quantum
  Hall ferromagnets,
  \href{https://doi.org/10.1103/PhysRevB.74.075423}{Phys. Rev. B {\bf
      74}, 075423 (2006).}
\bibitem{Lian17} Y. Lian and M. O. Goerbig, Spin-valley skyrmions in
  graphene at filling factor $\nu = -1$,
  \href{https://doi.org/10.1103/PhysRevB.95.245428}{Phys. Rev. B {\bf
      95}, 245428 (2017).}
\bibitem{jolicoeurQuantumHallSkyrmions2019} T. Jolicoeur and
  B. Pandey, Quantum Hall skyrmions at $\nu = 0,\pm 1$ in monolayer
  graphene,
  \href{https://doi.org/10.1103/PhysRevB.100.115422}{Phys. Rev. B {\bf
      100}, 115422 (2019).}
\bibitem{Zhou20} H. Zhou, H. Polshyn, T. Taniguchi, K. Watanabe, and
  A. F. Young, Solids of quantum Hall skyrmions in graphene,
  \href{https://doi.org/10.1038/s41567-019-0729-8}{Nat. Phys. {\bf
      16}, 154 (2020).}
\bibitem{Atteia21} J. Atteia and M. O. Goerbig, SU(4) spin wave in the
  $\nu = \pm 1$ quantum Hall ferromagnet in graphene,
  \href{https://doi.org/10.1103/PhysRevB.103.195413}{Phys. Rev. B {\bf
      103}, 195413 (2021).}
\bibitem{Pierce22} A. T. Pierce, Y. Xie, S. H. Lee, P. R. Forrester,
  D. S. Wei, K. Watanabe, T. Taniguchi, B. I. Halperin, and A. Yacoby,
  Thermodynamics of free and bound magnons in graphene,
  \href{https://doi.org/10.1038/s41567-021-01421-x}{Nat. Phys. {\bf
      18}, 37 (2022).}
\bibitem{Abanin06} D. A. Abanin, P. A. Lee, and L. S. Levitov,
  Spin-Filtered Edge States and Quantum Hall Effect in Graphene,
  \href{https://doi.org/10.1103/PhysRevLett.96.176803}{Phys. Rev. B
    {\bf 96}, 176803 (2006).}
\bibitem{kharitonov} M. Kharitonov, Phase diagram for the $\nu = 0$
  quantum Hall state in monolayer graphene,
  \href{https://doi.org/10.1103/PhysRevB.85.155439}{Phys. Rev. B {\bf
      85}, 155439 (2012).}
\bibitem{Young12} A. F. Young, C. R. Dean, L. Wang, H. Ren,
  P. Cadden-Zimansky, K. Watanabe, T. Taniguchi, J. Hone,
  K. L. Shepard, and P. Kim, Spin and valley quantum Hall
  antiferromagnetism in graphene,
  \href{https://doi.org/10.1038/nphys2307}{Nat. Phys. {\bf 8}, 550
    (2012).}
\bibitem{Young14} A. F. Young, J. D. Sanchez-Yamagishi, B. Hunt,
  S. H. Choi, K. Watanabe, T. Taniguchi, R. C. Ashoori, and
  P. Jarillo-Herrero, Tunable symmetry breaking and helical edge
  transport in a graphene quantum spin Hall state,
  \href{https://doi.org/10.1038/nature12800}{Nature {\bf 505}, 528
    (2014).}
\bibitem{Liu22} X. Liu, G. Farahi, C.-L. Chiu, Z. Papic, K. Watanabe,
  T. Taniguchi, M. P. Zaletel, and Ali Yazdani, Visualizing broken
  symmetry and topological defects in a quantum Hall ferromagnet,
  \href{https://doi.org/10.1126/science.abm3770}{Science {\bf 375},
    321 (2022).}
\bibitem{Cao18ins} Y. Cao, V. Fatemi, A. Demir, S. Fang,
  S. L. Tomarken, J. Y. Luo, J. D. Sanchez-Yamagishi, K. Watanabe,
  T. Taniguchi, E. Kaxiras, R. C. Ashoori, and P. Jarillo-Herrero,
  Correlated insulator behaviour at half-filling in magic-angle
  graphene superlattices
  \href{https://doi.org/10.1038/nature26154}{Nature {\bf 556}, 80
    (2018).}
\bibitem{Cao18sc} Y. Cao, V. Fatemi, S. Fang, K. Watanabe,
  T. Taniguchi, E. Kaxiras, and P. Jarillo-Herrero, Unconventional
  superconductivity in magic-angle graphene superlattices
  \href{https://doi.org/10.1038/nature26160}{Nature {\bf 556}, 43
    (2018).}
\bibitem{Yankowitz19} M. Yankowitz, S. Chen, H. Polshyn, Y. Zhang,
  K. Watanabe, T. Taniguchi, D. Graf, A. F. Young, and C. R. Dean,
  Tuning superconductivity in twisted bilayer graphene,
  \href{https://doi.org/10.1126/science.aav1910}{Science {\bf 363},
    1059 (2019).}
\bibitem{Lu20} X. Lu, P. Stepanov, W. Yang, M. Xie, M. A. Aamir,
  I. Das, C. Urgell, K. Watanabe, T. Taniguchi, G. Zhang, A. Bachtold,
  A. H. MacDonald, and D. K. Efetov, Superconductors, orbital magnets
  and correlated states in magic-angle bilayer graphene,
  \href{https://doi.org/10.1038/s41586-019-1695-0}{Nature {\bf 574},
    653 (2019).}
\bibitem{MacDonald19} For a review, see e.g.\ A. H. MacDonald, Bilayer
  graphene's wicked, twisted road,
  \href{https://doi.org/10.1103/Physics.12.12}{Physics {\bf 12}, 12
    (2019).}
\bibitem{Balents20} L. Balents, C. R. Dean, D. K. Efetov, and
  A. F. Young, Superconductivity and strong correlations in moir\'e
  flat bands
  \href{https://doi.org/10.1038/s41567-020-0906-9}{Nat. Phys. {\bf
      16}, 725 (2020).}
\bibitem{Ledwith21} P. J. Ledwith, E. Khalaf, and A. Vishwanath,
  Strong coupling theory of magic-angle graphene: A pedagogical
  introduction,
  \href{https://doi.org/10.1016/j.aop.2021.168646}{Ann. Phys. {\bf
      435}, 168646 (2021).}
\bibitem{Sharpe19} A. L. Sharpe, E. J. Fox, A. W. Barnard, J. Finney,
  K. Watanabe, T. Taniguchi, M. A. Kastner, and D. Goldhaber-Gordon
  Emergent ferromagnetism near three-quarters filling in twisted
  bilayer graphene,
  \href{https://doi.org/10.1126/science.aaw3780}{Science {\bf 365},
    605 (2019).}
\bibitem{Khalaf21} E. Khalaf, S. Chatterjee, N. Bultinck,
  M. P. Zaletel, and A. Vishwanath, Charged skyrmions and topological
  origin of superconductivity in magic-angle graphene,
  \href{https://doi.org/10.1126/sciadv.abf5299}{Sci. Adv. {\bf 7},
    eabf5299 (2021)}
\bibitem{weiElectricalGenerationDetection2018} D. S. Wei, T. van der
  Sar, S. H. Lee, K. Watanabe, T. Taniguchi, B. I. Halperin, and
  A. Yacoby, Electrical generation and detection of spin waves in a
  quantum Hall ferromagnet,
  \href{https://www.science.org/doi/epdf/10.1126/science.aar4061}{Science
    {\bf 362}, 229 (2018).}
\bibitem{pinczukSpectroscopicMeasurementLarge1992} A. Pinczuk, B. S.
  Dennis, D. Heiman, C. Kallan, L. Brey, C. Tejedor, S. Schmitt-Rink,
  L. N. Pfeiffer, K. W. West, Spectroscopic measurement of large
  exchange enhancement of a spin-polarized 2D electron gas,
  \href{https://doi.org/10.1103/PhysRevLett.68.3623}{Phys. Rev. Lett. {\bf
      68}, 3623 (1992).}
\bibitem{Stepanov18} P. Stepanov, S. Che, D. Shcherbakov, J. Yang,
  R. Chen, K. Thilahar, G. Voigt, M. W. Bockrath, D. Smirnov,
  K. Watanabe, T. Taniguchi, R. K. Lake, Y. Barlas, A. H. MacDonald,
  and C. N. Lau, Long-distance spin transport through a graphene
  quantum Hall antiferromagnet,
  \href{https://doi.org.10.1038/s41567-018-0161-5}{Nat. Phys. {\bf
      14}, 907 (2018).}
\bibitem{Chakraborty23} N. Chakraborty, R. Moessner, and B. Doucot,
  Magnon scattering off quantum Hall skyrmion crystals probes
  interplay of topology and symmetry breaking,
  \href{https://doi.org/10.1103/PhysRevB.108.104401}{Phys. Rev. B {\bf
      108}, 104401 (2023).}
\bibitem{Assouline21} A. Assouline, M. Jo, P. Brasseur, K. Watanabe,
  T. Taniguchi, Th. Jolicoeur, D. C. Glattli, N. Kumada, P. Roche,
  F. D. Parmentier, and P. Roulleau, Excitonic nature of magnons in a
  quantum Hall ferromagnet,
  \href{https://doi.org/10.1038/s41567-021-01411-z}{Nat. Phys. {\bf
      17}, 1369 (2021).}
\bibitem{iwasakiTheoryMagnonskyrmionScattering2014} J. Iwasaki,
  A. J. Beekman, and N. Nagaosa, Theory of magnon-skyrmion scattering
  in chiral magnets,
  \href{https://doi.org/10.1103/PhysRevB.89.064412}{Phys. Rev. B {\bf
      89}, 064412 (2014).}
\bibitem{garst} C. Sch\"utte and M. Garst, Magnon-skyrmion scattering
  in chiral magnets,
  \href{https://doi.org/10.1103/PhysRevB.90.094423}{Phys. Rev. B {\bf
      90}, 094423 (2014).}
\bibitem{liTowardsQuantumMagnonics2022} Z. Li, M. Ma, Z. Chen, K. Xie,
  and F. Ma, Interaction between magnon and skyrmion: Toward quantum
  magnonics, \href{https://doi.org/10.1063/5.0121314}{Journal of
    Applied Physics {\bf 132}, 210702 (2022).}
\bibitem{jiangSkyrmionsMagneticMultilayers2017} W. Jiang, G. Chen,
  K. Liu, J. Zang, S. G. E. te Velthuis, and A. Hoffmann, Skyrmions in
  magnetic multilayers,
  \href{https://doi.org/10.1016/j.physrep.2017.08.001}{Physics Reports
    {\bf 704}, 1 (2017).}
\bibitem{abl} G. Venkat, H. Fangohr, and A. Prabha, Absorbing boundary
  layers for spin wave micromagnetics,
  \href{https://doi.org/10.1016/j.jmmm.2017.06.057}{Journal of
    Magnetism and Magnetic Materials {\bf 450}, 34 (2018).}
\bibitem{Rajaraman82} R. Rajaraman, \emph{Solitons and Instantons}
  (Elsevier, Amsterdam, 1982).
\bibitem{Liu22spacing} X. Liu, J. I. A. Li, K. Watanabe, T. Taniguchi,
  J. Hone, B. I. Halperin, P. Kim, and C. R. Dean, Crossover between
  strongly coupled and weakly coupled exciton superfluids,
  \href{https://doi.org/10.1126/science.abg1110}{Science {\bf 375},
    205 (2022).}
\bibitem{sternTheoryInterlayerTunneling2001} A. Stern, S. M. Girvin,
  A. H. MacDonald, and N. Ma, Theory of Interlayer Tunneling in
  Bilayer Quantum Hall Ferromagnets,
  \href{https://doi.org/10.1103/PhysRevLett.86.1829}{Phys. Rev. Lett. {\bf
      86}, 1829 (2001).}
\bibitem{burkovPhaseTransitionSpin2002} A. Burkov, J. Schliemann,
  A. H. MacDonald, and S. M. Girvin, Phase transition and spin–wave
  dispersion in quantum Hall bilayers at filling factor $\nu = 1$,
  \href{https://doi.org/10.1016/S1386-9477(01)00299-5}{Physica E {\bf
      12}, 28 (2002).}
\bibitem{vanoosteromSolidAnglePlane1983} A. Van Oosterom and
  J. Strackee, IEEE Transactions on Biomedical Engineering,
  \href{http://ieeexplore.ieee.org/document/4121581/}{Physics Reports
    {\bf BME-30}, 125 (1983).}
\bibitem{ahoroni} A. Ahoroni, \emph{Introduction to the Theory of
    Ferromagnetism} (Oxford Science Publications, Oxford, England,
  2000).
\bibitem{Arovas99} D. P. Arovas, A. Karlhede, and D. Lillieh\"o\"ok,
  SU($N$) quantum Hall skyrmions,
  \href{https://doi.org/10.1103/PhysRevB.59.13147}{Phys. Rev. B {\bf
      59}, 13147 (1999).}
\bibitem{belavinMetastableStatesTwodimensional1975} A. A. Belavin
  and A. M. Polyakov, Metastable states of two-dimensional isotropic
  ferromagnets,
  Pis’ma Zh. Eksp. Teor. Fiz {\bf 22}, 503 (1975) [JETP Lett. {\bf
    22}, 245 (1975)].
\end{thebibliography}
\end{document}